\documentclass[%
 amsmath,amssymb,
 aps, physrev,
]{revtex4-2}

\usepackage{graphicx}
\usepackage{dcolumn}
\usepackage{bm}
\usepackage[number-unit-separator=~, range-units=single, separate-uncertainty = true]{siunitx}
\DeclareSIUnit[number-unit-product = \,]{\celsius}{\si{\degree}C}
\usepackage[labelformat=simple]{subcaption}

\usepackage{xcolor}
\usepackage{soul}

\usepackage{hyperref}
\definecolor{APSDarkBlue}{RGB}{45, 48, 146}
\hypersetup{
    colorlinks,
    linkcolor = APSDarkBlue,
    citecolor = APSDarkBlue,
    urlcolor = APSDarkBlue }
\usepackage[capitalise]{cleveref}

\makeatletter\def\Hy@Warning#1{}\makeatother 
\usepackage{microtype} 

\begin{document}


\title{Design and modeling of a liquid-lead dump concept for beamstrahlung radiation absorption in the CERN Future Circular e$^+$e$^-$ Collider}

\author{Silvio Candido}
\email{silvio.candido@cern.ch}
\author{Rui Franqueira Ximenes}
\email{rui.franqueira.ximenes@cern.ch}
\author{Anton Lechner}
\author{Alessandro Frasca}
\author{Giuseppe Lerner}
\author{Antonio Perillo Marcone}
\author{Regis Seidenbinder}
\author{Markus Widorski}
\author{Louise Jorat}
\author{Giacomo Lavezzari}
\author{Davide Bozzato}
\author{Gabriel Banks}
\author{Marco Calviani}
\affiliation{%
 CERN—European Organization for Nuclear Research, \\
 Geneva, CH-1211, Switzerland}%
\author{Luca Tricarico}
\author{Carlo Carrelli}
\author{Mariano Tarantino}
\affiliation{ENEA-Brasimone Research Center, 40032 Camugnano, Italy}

\date{\today}

\begin{abstract}
The electron-positron Future Circular Collider (FCC-ee) being developed at CERN will generate intense beamstrahlung radiation, thus requiring photon absorbers downstream of the interaction points. This work presents the conceptual design of flowing-liquid-lead absorbers capable of dissipating around 370 kW of photon power, with a mean photon energy of 62 MeV. Two configurations are investigated: an inclined-flow geometry, developed to increase the photon interaction length to maximize absorption, and a compact upstream slope with an additional pool, in which a free-surface lead flow intercepts the power peak before a downstream pool dissipates the remaining load. Photon-matter interactions are modeled using Monte Carlo simulations in \textsc{fluka}, while conjugate heat-transfer and free-surface dynamics are analyzed through computational-fluid-dynamics simulations using \textsc{ansys} \textsc{fluent}. Design refinements are introduced based on the simulated thermal and hydraulic performance and to mitigate secondary effects such as photon backscattering. Both configurations demonstrate stable operation within the 300-kg/s flow limit and maintain liquid-lead and structural temperatures within the operational range of 450--500 $^{\circ}$C. The results establish circulating liquid lead as a feasible and thermally robust baseline technology for beamstrahlung absorption in FCC-ee.
\end{abstract}

\maketitle



\section{\label{sec:1} Introduction}
\begin{figure}[!htb]
     \centering
     \includegraphics[width=0.9\textwidth]{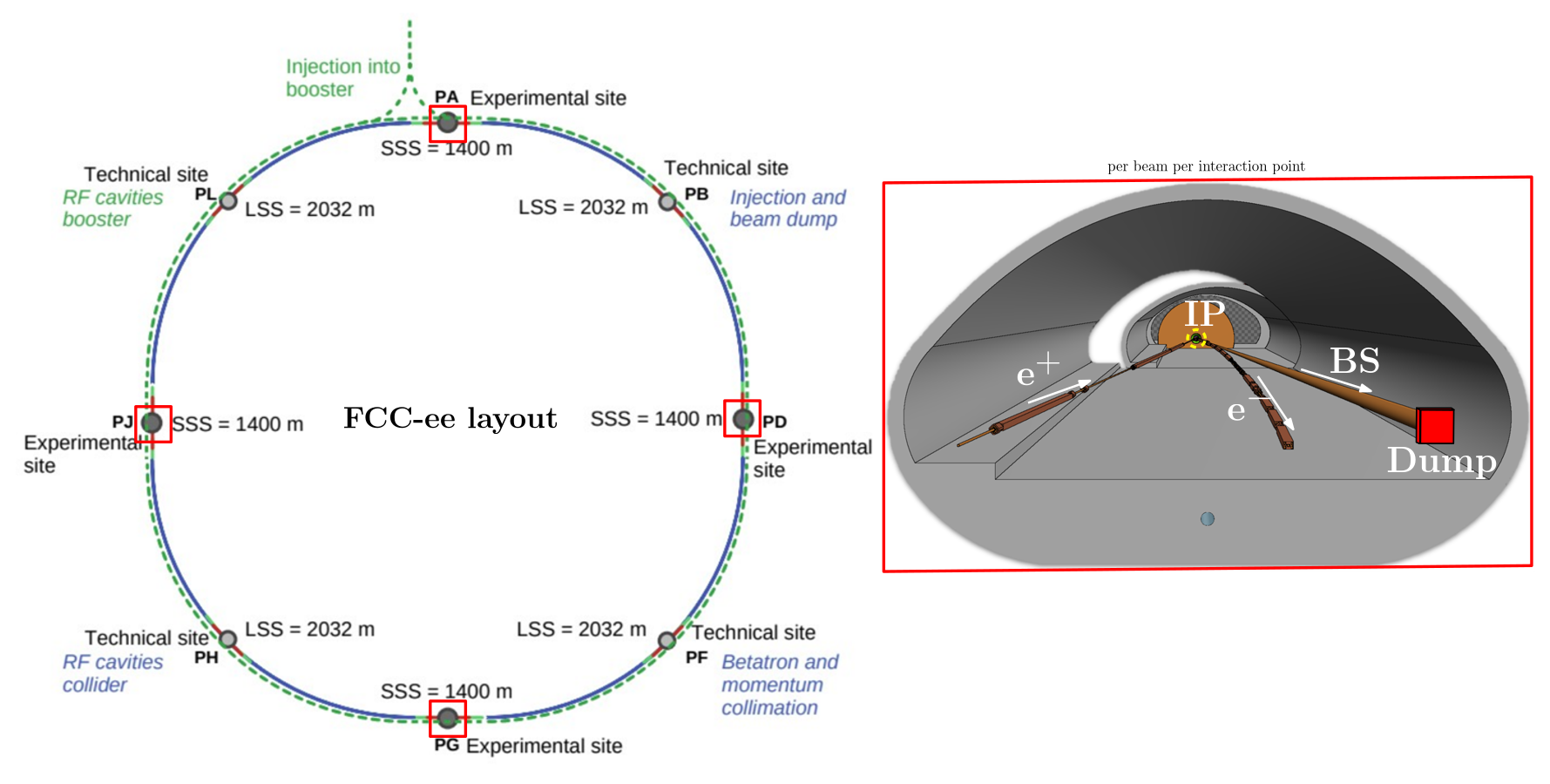}
     \caption{Diagram presenting the layout of FCC-ee, showing the four collision points (left)~\cite{Benedikt:2928793} and a schematic illustrating the position of the BS dump relative to each IP (right)~\cite{Frasca2024}.}
     \label{fig:1}
\end{figure}

The proposed electron-positron Future Circular Collider (FCC-ee) at CERN is designed to deliver unprecedented precision in the study of the Higgs boson, the electroweak sector, and the top quark~\cite{Abada2019, Blondel2022}. A direct consequence of its high-current operation is substantial beamstrahlung (BS) radiation; this is analogous to synchrotron radiation: during each bunch crossing, particles are deflected by the electromagnetic fields of the opposing bunch and emit highly collimated photons~\cite{Boscolo2023}. At the four FCC-ee interaction points (IPs), the BS power can reach about \SI{370}{\kilo\watt} during Z-pole operation, with mean photon energies of up to \SI{62}{\mega\electronvolt}~\cite{Boscolo2025}. These photon beams exit each IP and require dedicated absorbers downstream. With two dumps at each IP, a total of eight BS dumps are needed for the collider, as illustrated in \cref{fig:1}.

Effective management of BS radiation is required to protect downstream components and maintain stable collider operation under high-luminosity conditions~\cite{Piotrzkowski2023}. The photon power is sensitive to beam-beam parameters, and even small misalignments can increase the BS power by nearly a factor of~2~\cite{Boscolo2025}. A robust BS dump system capable of absorbing rapid power oscillations is therefore a prerequisite for the stable operation and long-term reliability of FCC-ee. The expected FCC-ee BS power emitted at each IP will reach several hundred kilowatts, concentrated in a narrow forward cone, a level unprecedented among previous lepton colliders~\cite{PhysRevAccelBeams.23.104402}. Unlike LEP or SuperKEKB, where the power was negligible and always dissipated in other beamline components, FCC-ee will require dedicated absorbers for each IP to safely manage these continuous photon loads. As a comparison, the SuperKEKB (e$^-$) IPs produce a BS beam with an average energy 63.3~times lower, and SuperKEKB (e$^+$) 23.2~times lower, than that expected for FCC-ee during Z-pole operation~\cite{Boscolo2023, DICARLO2022167453}.

Solid absorbers based on graphite or copper alloys face notable limitations in terms of melting, thermal fatigue~\cite{Sola2019}, cracking, and radiation damage~\cite{WAKAI2021152503}. As such, heavy liquid metals (HLMs) provide an attractive alternative absorber material~\cite{Pellemoine_2023, Calviani2024}. Due to their flowing nature, they allow efficient heat removal while avoiding the structural concerns of solid absorbers. Among candidate fluid metals, lead (Pb) combines high density and high photon-absorption efficiency with a relatively low melting point and a high boiling point. The ability to circulate liquid lead in a closed loop has been well demonstrated in facilities such as BULLET~\cite{Villanueva2025} and the Megawatt Pilot Experiment (MEGAPIE)~\cite{MEGAPIE2001}, meaning that it is a promising candidate material for FCC-ee. Thus, a liquid-lead-based design has been adopted as the FCC-ee baseline, although alternative solutions such as gas-cooled graphite discs are also under consideration~\cite{Boscolo2025}. A relevant precedent is the META:LIC target design~\cite{Class2014}, which demonstrated that lead-bismuth eutectic (LBE) circulation at 1.5--\SI{2}{\meter\per\second} can effectively manage high-power proton beams. Another notable example is the liquid-lithium target developed for the IFMIF-DONES facility~\cite{Nitti2025}, which successfully employs a continuously flowing film of lithium to withstand intense deuteron-beam power densities exceeding \SI{1}{\mega\watt\per\centi\meter\squared}. Together, these developments highlight the suitability and technological maturity of liquid-metal systems for demanding accelerator environments.

The proposed design uses pure lead rather than a eutectic Pb-Bi alloy. Although Pb-Bi mixtures have a lower melting point and favorable hydraulic behavior, the presence of bismuth can lead to the formation of volatile and radiotoxic \(^{210}\)Po under irradiation. In nuclear systems, the use of pure lead reduces \(^{210}\)Po production by about 4~orders of magnitude compared with LBE~\cite{Toshinsky2020}. Although this effect is less relevant for a photon absorber, secondary neutrons capable of triggering the production chain of \(^{210}\)Po~\cite{Zilges2022} are still generated. Moreover, the high cost of bismuth further favors the use of pure lead for a large-volume absorber.

The key design challenge for FCC-ee is to ensure sufficient interaction length of liquid lead to absorb the BS photon load while keeping the hydraulic system compact and practical for integration into the tunnel. Addressing this constraint requires geometries that maximize the effective photon path length while respecting hydraulic and integration limits.

In this work, we present the design and modeling of liquid-lead BS dump concepts, extending preliminary studies~\cite{candido2025ipac}. Two complementary approaches are explored: (i)~a sloped flow optimized with a double-slope geometry to suppress photon backscattering, and (ii)~a compact upstream slope combined with a back-pool configuration, in which a thin free-surface lead film intercepts the power peak before a downstream pool absorbs the remainder. Monte Carlo simulations with \textsc{fluka}~\cite{Battistoni2015, Ahdida2022} are used to quantify photon energy deposition and the resulting radiation environment, while computational fluid dynamics (CFD) simulations are used to resolve the coupled free-surface flow, heat transfer, and hydraulic constraints. Together, these tools establish a consistent framework for optimizing the absorber geometry, ensuring safe operation, and demonstrating the feasibility of liquid lead as a baseline technology for BS absorption in FCC-ee.

The remainder of this paper is organized as follows. Section~\ref{sec:2} introduces the properties of BS radiation and liquid-lead systems. Section~\ref{sec:3} outlines the numerical methodology, including the CFD and \textsc{fluka} setups. Section~\ref{sec:4} presents the results, focusing on absorption efficiency, flow behavior, and backscattering mitigation. Section~\ref{sec:5} concludes the paper, providing a summary of the findings and outlining directions for further design optimization.

\section{Liquid-lead beamstrahlung dump concept}\label{sec:2}

\subsection{Properties of beamstrahlung radiation}
During each bunch crossing at the IP of a collider, particles are deflected by the electromagnetic field of the opposing bunch. The resulting beam-beam kick (electromagnetic deflection) bends their trajectories and induces photon emission, known as beamstrahlung. This process is particularly relevant in lepton colliders, and its characteristics depend on several factors, including the beam energy $E$, the bunch intensity $N_{\mathrm{e}}$, and the bunch shape at collision.

In FCC-ee, the large bunch population and small transverse beam dimensions are expected to produce strong BS emission (high photon flux and elevated average photon energy). Table~\ref{tab:table2a} summarizes the main FCC-ee beam parameters, along with the properties of the corresponding BS radiation as calculated using \textsc{guinea-pig\footnotesize{++}} and presented in Ref.~\cite{Ciarma:2023iat} for optics V22. In the absence of longitudinal shifts or transverse offsets between the colliding bunches, each beam radiates about \SI{370}{\kilo\watt} of BS power at the Z pole and about \SI{77}{\kilo\watt} during ${\rm t}\bar{\rm t}$ operation. Compared with more recent optics versions, V22 assumes a higher bunch population, leading to a significantly higher BS power. Considering optics V22 therefore provides conservative estimates. Deviation from nominal conditions can strongly affect BS emission. In particular, the BS power is highly sensitive to bunch separation at collision. Due to the flat beams of FCC-ee, a vertical offset of only a few tens of nanometers can increase the radiated power by more than 50\%.

The emission of BS photons is confined to a very narrow cone, with an opening angle that is inversely proportional to the beam energy. Consequently, two intense photon beams emerge collinear with the outgoing beams from each IP.

\begin{table}[!htb]
\caption{\label{tab:table2a}FCC-ee beam parameters for optics V22 and associated BS radiation properties simulated with \textsc{guinea-pig\footnotesize{++}}~\cite{Ciarma:2023iat} at the Z pole and in ${\rm t}\bar{\rm t}$ operation.}
\begin{ruledtabular}
\begin{tabular}{lcc}
& Z pole & ${\rm t}\bar{\rm t}$ \\
\hline Beam energy, $E$ (\si{\giga\electronvolt}) & 45.6 & 182.5 \\
Bunches/beam, $n_{\mathrm{b}}$ & \num{10000} &  36 \\
Bunch intensity, $N_{\mathrm{e}}$ ($10^{11}$) & 2.43 & 2.64 \\
BS mean photon energy (\si{\mega\electronvolt}) & 1.7 & 62.3 \\
BS power (one beam) (\si{\kilo\watt}) & 370 & 77 \\
BS beam horizontal divergence, $\sigma_{px}$ (\SI{}{\micro\radian}) & 91.8 & 44.6 \\
BS beam vertical divergence, $\sigma_{py}$ (\SI{}{\micro\radian}) & 49.2 & 50.3 \\
BS beam horizontal spot size center @ \SI{500}{\meter} (\si{\centi\meter}) & 2.27 & 0.473 \\
BS beam horizontal spot size @ \SI{500}{\meter}, $\sigma_x^{500\,\mathrm{m}}$ (\si{\centi\meter}) & 4.59 & 2.23 \\
BS beam vertical spot size @ \SI{500}{\meter}, $\sigma_y^{500\,\mathrm{m}}$ (\si{\centi\meter}) & 2.46 & 2.51
\end{tabular}
\end{ruledtabular}
\end{table}

\subsection{Design requirements and technical constraints}

In the current baseline~\cite{Benedikt:2928793}, the BS dump is placed \SI{500}{\meter} downstream of the IP. This ensures sufficient separation between the beamlines and the dump system while also providing the benefit of a reduced peak power density due to the larger photon-beam size and preventing unwanted background to the detectors. BS radiation is directed toward the BS dump through a dedicated extraction line branching from the beamline outgoing from the IP. Due to the beam-beam kicks experienced by the radiating particles at the IP, the emitted photons have, on average, a nonzero horizontal emission angle with respect to the outgoing beam axis. This effect, which is more pronounced in Z mode because of the lower magnetic rigidity of the beam particles, leads to a horizontal offset of a few centimeters between the center of the beam spot and the center of the dump (Table~\ref{tab:table2a}).

In this configuration, additional radiation produced collinearly with the outgoing beam is intercepted by the dump. This includes synchrotron radiation emitted in the downstream dipoles and in the final-focus magnets, as well as photons produced within the detector magnetic fields~\cite{Boscolo2017}. Although these sources generate several hundred kilowatts of synchrotron radiation in total within the interaction region, estimates of power flow in the interaction region indicate that only a very small fraction propagates downstream and reaches the vicinity of the BS dump~\cite{Boscolo2026}. The Z pole is the dominant contribution driving the absorber design; a dump optimized for Z-pole operation will also be capable of handling the photon spectrum in ${\rm t}\bar{\rm t}$ operation.

The high-energy tail of the BS spectrum (hundreds of \si{\mega\electronvolt} at the Z pole and a few \si{\giga\electronvolt} during ${\rm t}\bar{\rm t}$ operation) triggers electromagnetic cascades that develop over several radiation lengths~$X_0$. Because $X_0\propto A/(Z^2\rho)$, lead---with its short $X_0\approx \SI{0.56}{\centi\meter}$---provides efficient absorption in a compact volume. Low-$Z$ materials such as graphite ($X_0\approx \SI{20}{\centi\meter}$) would require substantially larger structures~\cite{PDG2024}. The absorber must also tolerate the highly localized energy deposition from the photon shower. An enclosed solid absorber would experience large thermal gradients, creating stresses that could lead to deformation or cracking. A flowing liquid absorber with a free surface circumvents these limitations by continuously renewing the interaction region and carrying away the deposited heat.

A preliminary \textsc{fluka} study~\cite{Frasca2024} modeled a \SI[number-unit-separator=\text{-}]{20}{\centi\meter}-thick, $70 \times \SI[number-unit-separator=\text{-}]{70}{\centi\meter\squared}$ lead volume and showed absorption above 90\% of the BS power at the Z pole; however, such a thickness cannot be realized as a free-falling film: sustaining a \SI[number-unit-separator=\text{-}]{20}{\centi\meter}-thick curtain over a \SI[number-unit-separator=\text{-}]{70}{\centi\meter}-wide footprint would require a velocity $v=\sqrt{2gh}\approx\SI{3.7}{\meter\per\second}$ and a mass flow of $\dot{m}\approx \SI{5890}{\kilogram\per\second}$ (for $\rho_{\rm Pb}=\SI{11340}{\kilogram\per\meter\cubed}$). This exceeds the feasible hydraulic range by more than an order of magnitude.

In practice, the circulating liquid-lead system is limited to $\dot{m}_{\max}\approx 300$--\SI{350}{\kilogram\per\second}, as determined by pump capability, allowable pressure drops, and pipe diameters (typically DN~150--DN~200). Operating near this limit already implies velocities of $\mathcal{O}(1)$~\si{\meter\per\second} and pressure drops of 0.1--\SI{0.3}{\mega\pascal} over tens of meters. Higher flow rates would require substantially larger pipes, heavier shielding, and more powerful pumps, all of which are incompatible with tunnel-space constraints.

\begin{figure}[!htb]
    \centering
    \begin{subfigure}[b]{0.45\linewidth}
        \includegraphics[width=\linewidth]{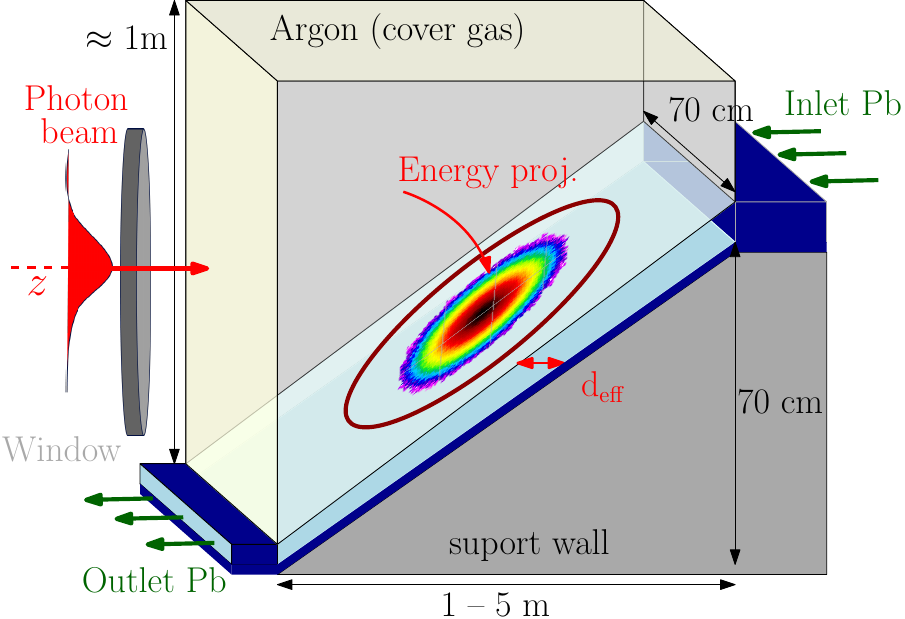}
        \caption{}
        \label{fig:2a}
    \end{subfigure}
    \begin{subfigure}[b]{0.45\linewidth}
        \includegraphics[width=\linewidth]{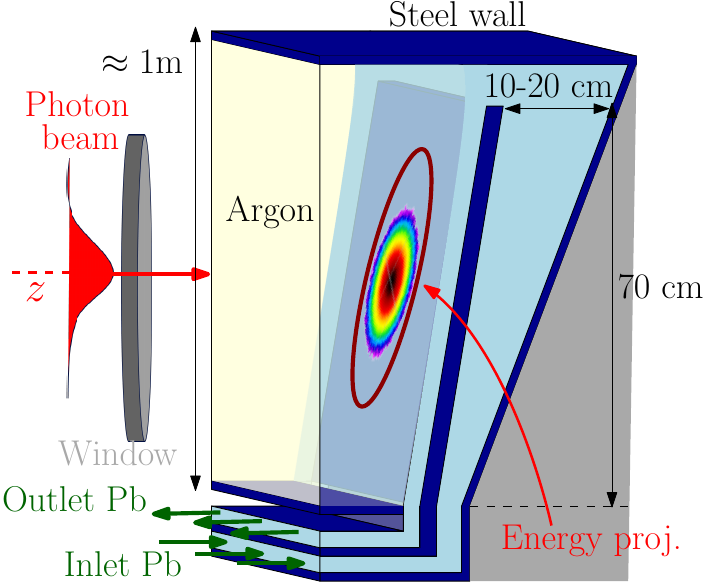}
        \caption{}
        \label{fig:2b}
    \end{subfigure}
    \caption{Schematics for the two main concepts of liquid lead flow for the BS dump: (a)~slope concept; (b)~upstream curtain/pool concept.}
    \label{fig:2}
\end{figure}

\Cref{fig:2} shows schematics of the two proposed concepts. The slope concept [\cref{fig:2a}] relies on full absorption of the photon energy within a sloped flowing lead assembly, requiring a long dump, approximately \SI{5}{\meter}, to achieve the necessary effective absorption thickness. The upstream curtain/pool concept [\cref{fig:2b}] seeks to absorb the initial peak power of the photon beam with a thin, free-surface lead curtain, and this is followed by a lead pool for absorbing the residual power. This design is intended to make the dump more compact, approximately \SI{1}{\meter} long, requiring less space for integration into the tunnel and reducing the volume of shielding needed to protect the surrounding equipment and personnel.

The main challenge in this work is to design these systems to maximize the effective thickness while maintaining a feasible flow rate of liquid lead, with minimal impact on integration into the FCC tunnel. The effective thickness, \(d_{\mathrm{eff}}\), is defined as the straight-line distance that the photon beam travels inside the liquid lead. Formally, if \(\Gamma\) is the beam path, then \(d_{\mathrm{eff}}=\int_{\Gamma}\chi_{m}(\mathbf{x})\,\mathrm{d}z\), where the subscript $m\in\{\mathrm{Pb},\mathrm{Ar},\mathrm{SS}\}$ in $\chi_m(\mathbf{x})$ indicates the material (with ``SS'' indicating stainless steel).

The liquid-lead circuit requires an inert cover gas to prevent oxidation of the free surface. Argon is typically used because of its chemical inertness, availability, and higher molecular weight than helium, which aids containment. Maintaining a sufficiently low oxygen partial pressure is essential to minimize oxide formation~\cite{Handbook}.

\section{Mathematical formulation and numerical methods}\label{sec:3}

\subsection{Governing equations for computational-fluid-dynamics simulations}

The governing equations for the two-phase flow in the computational domain $\Omega_0 \subset \mathbb{R}^3$ are those for the conservation of mass (continuity equation), momentum (Navier--Stokes equations), and energy.

For a two-phase flow in the computational domain $\Omega_0 \subset \mathbb{R}^3$ the conservation equations for mass, momentum, and energy are written as
\begin{equation}
\frac{\partial \rho}{\partial t} + \nabla \cdot (\rho \mathbf{u}) = 0, \quad \text{in} \quad \Omega_0 \times [0, t],
\end{equation}
\begin{equation}
\frac{\partial (\rho \mathbf{u})}{\partial t} + \nabla \cdot (\rho \mathbf{u} \mathbf{u}) = -\nabla p + \nabla \cdot \left( \mu \nabla \mathbf{u} \right) + \rho \mathbf{g} + \mathbf{f}_\gamma, \quad \text{in} \quad \Omega_0 \times [0, t],
\end{equation}
\begin{equation}
\frac{\partial (\rho E)}{\partial t} + \nabla \cdot \left( \mathbf{u} (\rho E + p) \right) = \nabla \cdot \left( \kappa_{\text{eff}} \nabla T \right) + S_{\mathrm{h}}, \quad \text{in} \quad \Omega_0 \times [0, t],
\end{equation}
respectively, where $\rho$ is the fluid density, $t$ is time, $\mathbf{u}$ is the velocity vector, $p$ is the pressure, $\mu$ is the dynamic viscosity, $\mathbf{g}$ is the gravitational acceleration vector, and $\mathbf{f}_{\gamma}$ represents the surface-tension force at the liquid-gas interface. The variable $E$ denotes the total specific energy, defined as the sum of internal and kinetic energy per unit mass. The term $\kappa_{\mathrm{eff}}$ is the effective thermal conductivity, $T$ is the temperature, and $S_{\mathrm{h}}$ is a volumetric heat source representing the power deposited by the photon beam in the material domain.

The volume-of-fluid (VoF) method~\cite{ansys_fluent} is used to solve the motion of two immiscible phases (liquid lead and cover gas) by tracking the liquid volume fraction $\alpha$ in each computational cell. This scalar is transported according to the advection equation
\begin{equation}
\frac{\partial \alpha}{\partial t} + \nabla \cdot (\alpha\,\mathbf{u}) = 0,
\end{equation}
where $\alpha$ denotes the local cell volume fraction of liquid lead. In each cell, mixture properties are obtained by volume-fraction weighting, which is important to maintain the consistency of mass, momentum, and energy transfer across the interface~\cite{Candido2023},
\begin{equation}
\phi = \alpha\,\phi_{\mathrm{Pb}} + (1 - \alpha)\,\phi_{\mathrm{Ar}}, \qquad \phi \in \{\rho, \mu, k, c_p\},
\end{equation}
where $\rho$, $\mu$, $k$, and $c_p$ denote the density, dynamic viscosity, thermal conductivity, and specific heat capacity, respectively, and the subscripts $\mathrm{Pb}$ and $\mathrm{Ar}$ refer to liquid lead and argon, respectively. The thermophysical properties of liquid lead are specified as functions of temperature~\cite{Handbook}, while the argon cover gas is treated as incompressible with properties assumed constant and evaluated at a reference temperature of \SI{400}{\celsius}. The main correlations used are summarized in Table~\ref{tab:table2}. Surface tension is included to account for liquid-gas interfacial forces.

\begin{table}[!htb]
\caption{\label{tab:table2} Thermophysical properties of liquid Pb~\cite{Handbook} and Argon~\cite{NIST_SRD69_Fluids}.}
\begin{ruledtabular}
\begin{tabular}{lcc}
Property (per phase $m$ for $T \in [600,1200]$~K) & Pb & Ar \\
\hline Mass density, $\rho_{m}$ (\si{\kilogram\per\meter\cubed}) & $11\,441 - 1.2795\,T$ & $1.6228$ \\
Dynamic viscosity, $\eta_{m}$ (\si{\pascal\second}) & $4.55\times10^{-4}\,\exp(1069/T)$ & $2.125\times10^{-5}$ \\
Thermal conductivity, $k_{m}$ [\si{\watt\per\meter\per\kelvin}] & $9.2 + 0.011\,T$ & $0.030$ \\
Heat capacity, $c_{p(m)}$ [\si{\joule\per\kilogram\per\kelvin}]\footnotemark[1] & $a_0 + a_1 T + a_2 T^2 + a_{-2}T^{-2}$ & $540$ \\
Surface tension, $\sigma$ (\si{\newton\per\meter}) & $(525.9 - 0.113\,T)\times10^{-3}$ & -- \\
\end{tabular}
\end{ruledtabular}
\footnotetext[1]{For Pb heat capacity: $a_0=176.2$, $a_1=-4.923\times10^{-2}$, $a_2=1.544\times10^{-5}$, $a_{-2}=-1.524\times10^{6}$.}
\end{table}

Near-wall turbulence and potential flow separation are modeled using the shear-stress transport $k$--$\omega$ formulation~\cite{MENTER1994} with automatic near-wall treatment. Because liquid lead has a very low molecular Prandtl number ($\mathrm{Pr} \approx 0.03$), heat diffuses much faster than momentum, and the classical Reynolds analogy does not hold. In Reynolds-averaged Navier--Stokes simulations, this discrepancy is accounted for through the turbulent Prandtl number $\mathrm{Pr}_{\mathrm{t}}$, which governs the ratio between turbulent momentum and heat transport via $\alpha_{\mathrm{t}} = \nu_{\mathrm{t}}/\mathrm{Pr}_{\mathrm{t}}$. The conventional choice $\mathrm{Pr}_{\mathrm{t}} \approx 0.85$ is known to produce significant errors for HLM flows~\cite{ChengTak2006}. Following recent assessments of turbulent heat transfer in lead and LBE, an effective turbulent Prandtl number of $\mathrm{Pr}_{\mathrm{t}} = 2.5$ is adopted, consistent with correlations proposed for HLM flows~\cite{Shen2021}. Given the thin-film nature of the flow and the use of wall functions, the first off-wall node is positioned within the logarithmic layer ($y^+ \approx 30$--$100$). For $\mathrm{Pr} \ll 1$, the thermal boundary layer is much thicker than the viscous sublayer, ensuring that this first cell remains within the thermal layer, and the temperature wall function provides an adequate wall heat-flux estimate for the conjugate heat-transfer calculation.

\subsection{Beamstrahlung energy-deposition field}

The volumetric power-density fields used in the thermal-hydraulic simulations are obtained with \textsc{fluka}~\cite{Battistoni2015, Ahdida2022}. The BS photon sample, generated with \textsc{guinea-pig\footnotesize{++}}~\cite{Ciarma:2023iat}, is injected at the IP and transported through a simplified model of the FCC-ee interaction region and dump geometry~\cite{Frasca2024}. \textsc{fluka} then computes the spatial distribution of deposited power, $Q_m(\mathbf{x})$, for all relevant materials $m\in\{\mathrm{Pb},\mathrm{Ar},\mathrm{SS}\}$.

To couple the radiation transport to the CFD solution, the \textsc{fluka}-scored volumetric power density is incorporated into the energy equation through the source term
\begin{equation}
    S_{\mathrm{h}}(\mathbf{x})=
    \sum_{m\in\{\mathrm{Pb},\mathrm{Ar},\mathrm{SS}\}}
    \chi_m(\mathbf{x})\,Q_m(\mathbf{x}).
\end{equation}
In the liquid-gas region, the material indicator is mapped using the VoF volume fractions, with $\chi_{\mathrm{Pb}}=\alpha$ and $\chi_{\mathrm{Ar}}=1-\alpha$. In solid regions, the deposition is applied directly based on the local material. This procedure ensures a conservative and consistent indirect coupling between photon-matter interactions and the transient thermal-hydraulic response of the system.

\subsection{Hydraulic system considerations}

The hydraulic system driving the liquid-lead flow imposes important constraints on the design of the BS absorber. In particular, the total mass flow rate must remain below approximately \SI{300}{\kilogram\per\second} to ensure mechanical feasibility, limit pump-power requirements, and maintain integration compatibility. Exceeding this flow rate would require substantial increases in pipe diameter, pumping head, and heat-exchanger capacity, ultimately impacting reliability and cost.

As an order-of-magnitude estimate, at $\dot{m}=\SI{300}{\kilogram\per\second}$ with a characteristic line velocity $u \approx \SI{1}{\meter\per\second}$, in DN~150--DN~200 lines, the estimated pressure drop over approximately $\SIrange{50}{100}{\meter}$ of the primary circuit is $\Delta p\approx 0.07\text{--}\SI{0.25}{\mega\pascal}$, to which static elevation adds $\rho g\Delta z\simeq\SI{0.10}{\mega\pascal}$ per meter. The resulting hydraulic power is $P_{\mathrm{h}}=\mathcal{O}(\SI{10}{\kilo\watt})$. Electrical input depends on pump type: roughly $14\text{--}\SI{20}{\kilo\watt}$ for centrifugal pumps ($\eta\approx0.5$--$0.7$) and $33\text{--}\SI{100}{\kilo\watt}$ for electromagnetic pumps ($\eta\approx0.1$--$0.3$).

Given these limitations, to achieve the desired effective photon absorption within the flow-rate limit, optimization becomes critical. The absorber geometry should therefore be tailored to maximize the projected interaction length of the beam while respecting a fixed hydraulic power. This strategy enables the system to meet the BS power-absorption requirements---approximately \SI{370}{\kilo\watt} per dump---without exceeding allowable hydraulic parameters. Comparable liquid-lead circulation loops, such as the BULLET facility~\cite{Villanueva2025}, operate at temperatures up to \SI{480}{\celsius} and velocities up to \SI{6}{\meter\per\second}, providing a useful benchmark for feasible flow and thermal limits in engineered lead systems.

\subsection{Special considerations regarding corrosion and erosion}

The liquid-lead absorber will operate near \SI{400}{\celsius} with flow velocities up to \SI{1}{\meter\per\second}---conditions approaching the thresholds at which corrosion and oxidation become critical for HLMs. Ensuring chemical compatibility between the circulating lead and the walls of the steel vessel is therefore a key consideration for the long-term reliability of the dump.

In HLM systems, material degradation occurs primarily through dissolution, oxidation, and erosion. Dissolution transfers iron from the steel into the liquid metal and is accelerated by high temperatures and flow-induced mixing~\cite{ZHANG2008351}. Oxidation depends on the dissolved oxygen concentration, which must be maintained within the steel passivation range, $C_{\mathrm{O},\mathrm{m}}<C_{\mathrm{O}}<C_{\mathrm{O},\mathrm{sat}}$, to form a stable protective oxide without generating lead oxides~\cite{Bassini2017}. For liquid lead at \SI{400}{\celsius}, this corresponds to $C_{\mathrm{O}} \approx 10^{-10}$ to $10^{-4}$~wt\,\%. Passivation deteriorates above \SI{480}{\celsius} and significantly at \SI{500}{\celsius}, with the latter defining the upper temperature limit for the walls of the system. Erosion and flow-accelerated corrosion result from turbulent shear at the metal-steel interface~\cite{Handbook}. Experiments indicate that for flowing Pb, corrosion remains modest below velocities of \SI{1}{\meter\per\second} but increases sharply above \SI{2}{\meter\per\second} as shear removes the oxide film~\cite{Handbook,Bassini2017}. These findings directly informed the hydraulic design of the dump, which limits the mean flow velocity to below \SI{1}{\meter\per\second} and maintains bulk temperatures under \SI{500}{\celsius}.

To control the oxygen potential, the system operates under an argon cover gas and incorporates purification loops similar to those used in existing lead circuits. In summary, keeping the oxygen concentration within the passivation range, the temperature below \SI{480}{\celsius}, and the flow velocity under \SI{1}{\meter\per\second} ensures that corrosion and erosion remain negligible throughout the absorber's lifetime.

\section{Results and discussion}\label{sec:4}

This section outlines the key engineering considerations and conceptualizations, and it presents the solutions implemented to address the problem.

\subsection{Optimizing the liquid-lead slope for effective thickness}

If we consider a free-falling curtain (neglecting dissipative losses), injected vertically through a $70 \times \SI[number-unit-separator=\text{-}]{20}{\centi\meter\squared}$ nozzle with a mass flow rate of \SI{300}{\kilogram\per\second}, this accelerates under gravity and thins according to mass conservation. Using a ballistic velocity increase and a uniform thinning across the \SI[number-unit-separator=\text{-}]{70}{\centi\meter} width, we estimate a thickness at $z=\SI{35}{\centi\meter}$ (center) of approximately \SI{1.09}{\centi\meter}. If one side is assumed to be a nonslip wall, an asymmetric profile develops, yielding a centerline thickness of \SI{2.02}{\centi\meter}.

For the present concept of a liquid-lead dump for BS radiation, an effective thickness of approximately \SI{20}{\centi\meter} at the beam center is required to absorb more than 90\% of the beam energy; thus, a simple curtain cannot be used. To achieve our goal without increasing the hydraulic demands on the system or increasing its dimensions, a sloped flow of liquid lead with a maximum length of \SI{5}{\meter} was proposed. With an inclined flow surface, the photon beam traverses a larger material depth due to the projection of its path through the liquid.

\begin{figure}[!htb]
     \centering
     \begin{subfigure}[b]{0.5\linewidth}
         \centering
         \includegraphics[width=\linewidth]{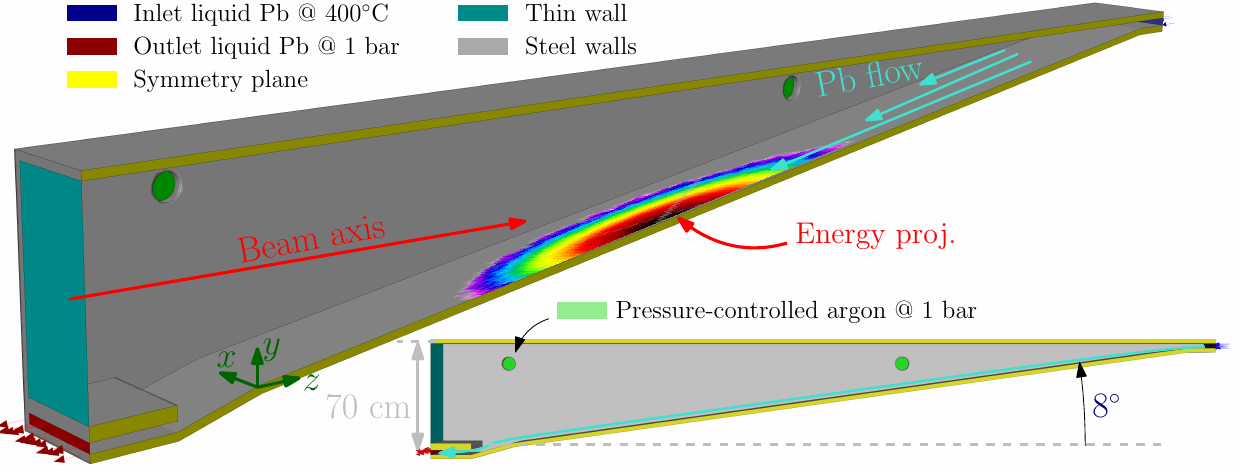}
         \caption{}
         \label{fig:3a}
     \end{subfigure}
     \hspace{-0.025\linewidth}
     \begin{subfigure}[b]{0.5\linewidth}
         \centering
         \includegraphics[width=\linewidth]{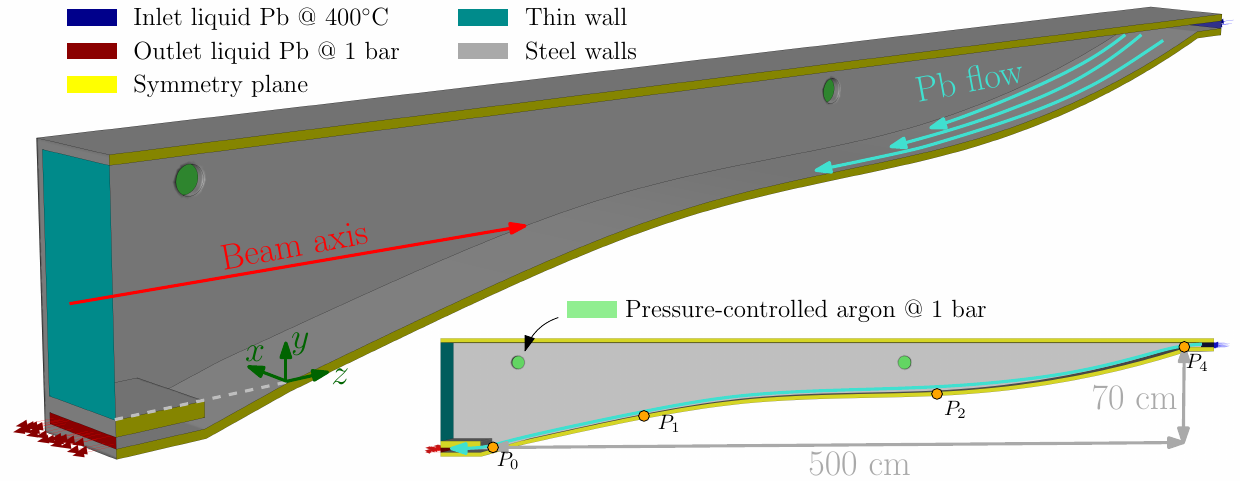}
         \caption{}
         \label{fig:3b}
     \end{subfigure}
     \caption{Schematics of the two slope designs for the liquid-lead flow in the BS dump, showing the (a)~linear (L-slope) and (b)~optimized curved (S-slope) shapes.}
     \label{fig:3}
\end{figure}

Initially, a linear slope (L-slope) inclined at $8^{\circ}$ was considered, and a schematic of this is presented in~\cref{fig:3a}. With this shallow angle, the effective path through the liquid lead increases as $t_{\mathrm{eff}}=t/\sin\alpha$, corresponding to an effective thickness of about \SI{7.2}{\centi\meter} for a \SI[number-unit-separator=\text{-}]{1}{\centi\meter}-thick film. Subsequently, an optimized curved profile (S-slope) was introduced, as illustrated in \cref{fig:3b}. Both configurations share identical boundary conditions, and the only difference lies in the slope geometry.

The S-slope geometry was optimized by defining an S-shaped profile using cubic spline interpolation, which is well suited to achieving smooth transitions between control points. The spline is constructed as a piecewise cubic polynomial passing through a set of control points $(z_i, y_i)$. In this case, the curve is defined by four points, $P_0 = (0, 0)$, $P_1 = (1.66, 0.31)$, $P_2 = (3.32, 0.38)$, and $P_3 = (0, 0.7)$. This fit yields a profile that maximizes the effective thickness near the center of the photon beam (where the energy deposition is highest) while maintaining a continuous transition along the length of the absorber.

\begin{figure}[!htb]
     \centering
     \begin{subfigure}[b]{0.49\linewidth}
         \centering
         \includegraphics[width=\textwidth]{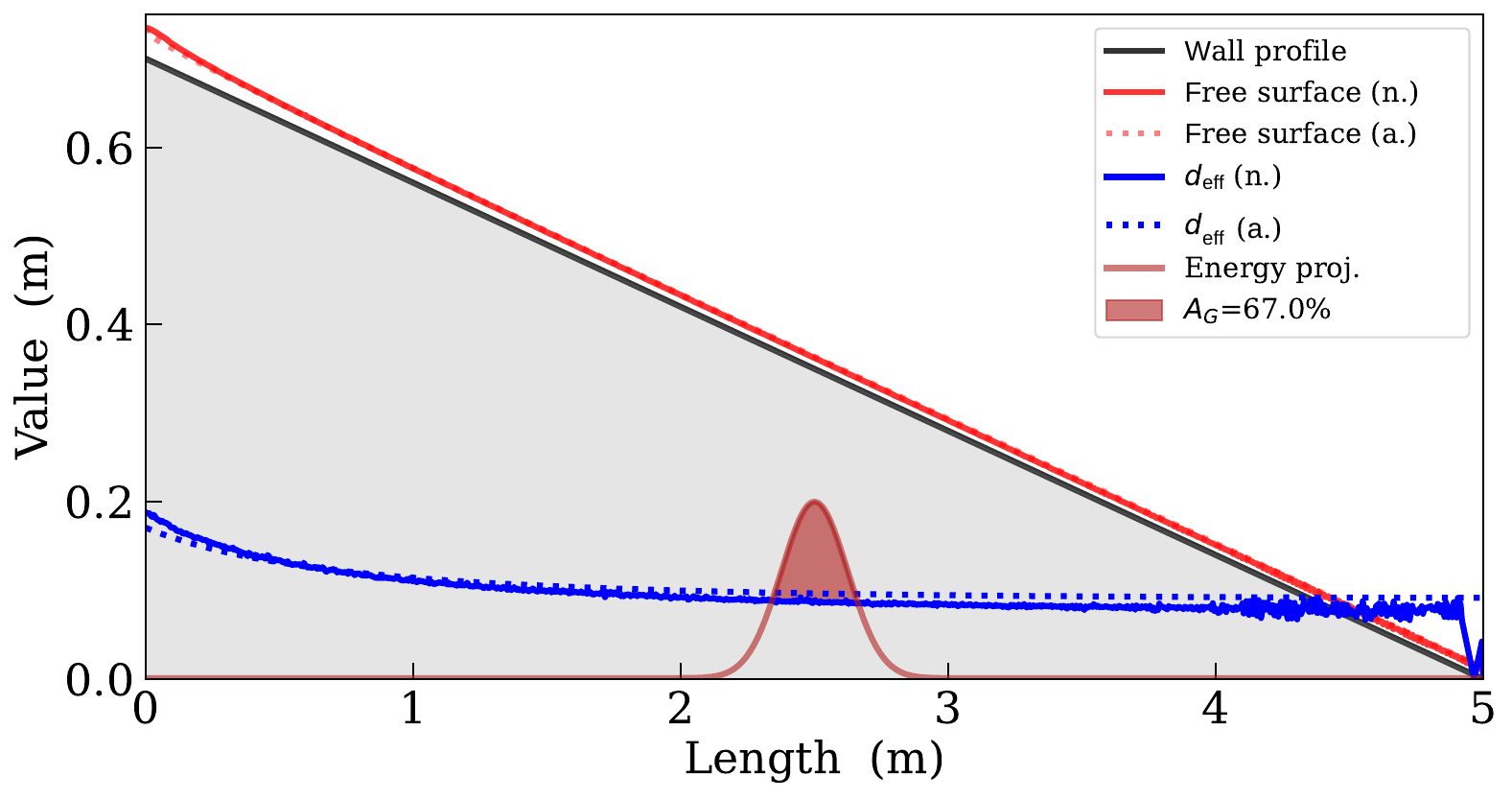}
         \caption{}
         \label{fig:ap1a}
     \end{subfigure}
     \hfill
     \begin{subfigure}[b]{0.49\linewidth}
         \centering
         \includegraphics[width=\textwidth]{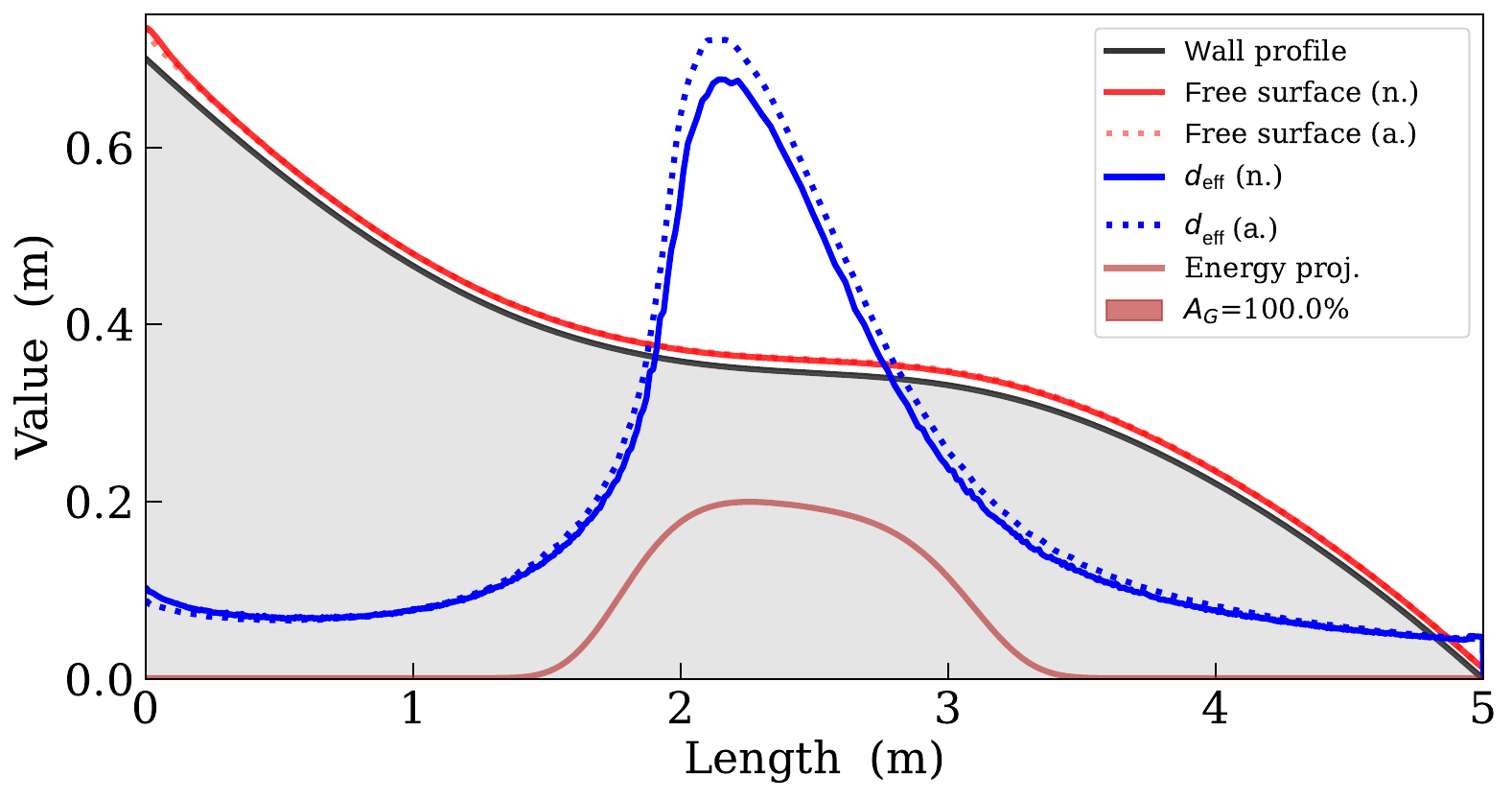}
         \caption{}
         \label{fig:ap1b}
     \end{subfigure}
     \hfill
        \caption{Free-surface profiles for the (a)~L-slope and (b)~S-slope with the corresponding effective thickness predicted by the analytical and CFD models. Definitions: n., numerical results; a., analytical calculations.}
        \label{fig:ap1}
\end{figure}

\Cref{fig:ap1} shows a comparison of the effective thickness ($d_\mathrm{eff}$) along the free surface of the flow between the CFD results and the analytical predictions. It can be seen that there is a discrepancy of less than 2\%. It is worth noting that in the final meter of the linear case, the CFD results capture wave instabilities that the analytical model cannot resolve. Analytical predictions based on the Colebrook--White friction model for free-surface flows~\cite{Bellos2018a} were used to validate the free-surface calculations (see Appendix~\ref{app:A}).

The L-slope produces a gradual acceleration of the liquid lead, resulting in a narrow high-velocity region and thinner coverage downstream. In contrast, the S-slope redistributes the flow and generates a broader high-velocity band, supporting a consistently greater effective thickness along the photon-impact path. Regions of low $d_{\text{eff}}$ evident in the linear case are effectively suppressed by the S-slope, ensuring that the photon beam traverses a thicker and more uniform absorber layer. These simplified two-dimensional results anticipate the trends observed in the full three-dimensional CFD analysis, in which the S-slope demonstrates superior absorption capacity and thermal performance.

\begin{figure}[!htb]
     \centering
     \begin{subfigure}[b]{0.8\linewidth}
         \centering
         \includegraphics[width=\textwidth]{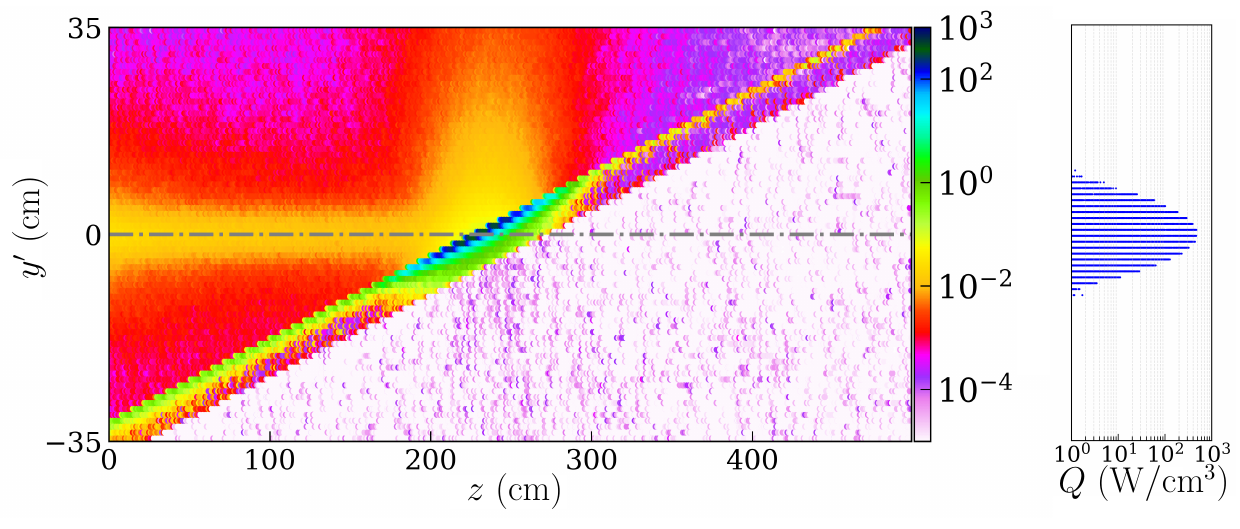}
         \caption{}
         \label{fig:5a}
     \end{subfigure}
     \begin{subfigure}[b]{0.8\linewidth}
         \centering
         \includegraphics[width=\textwidth]{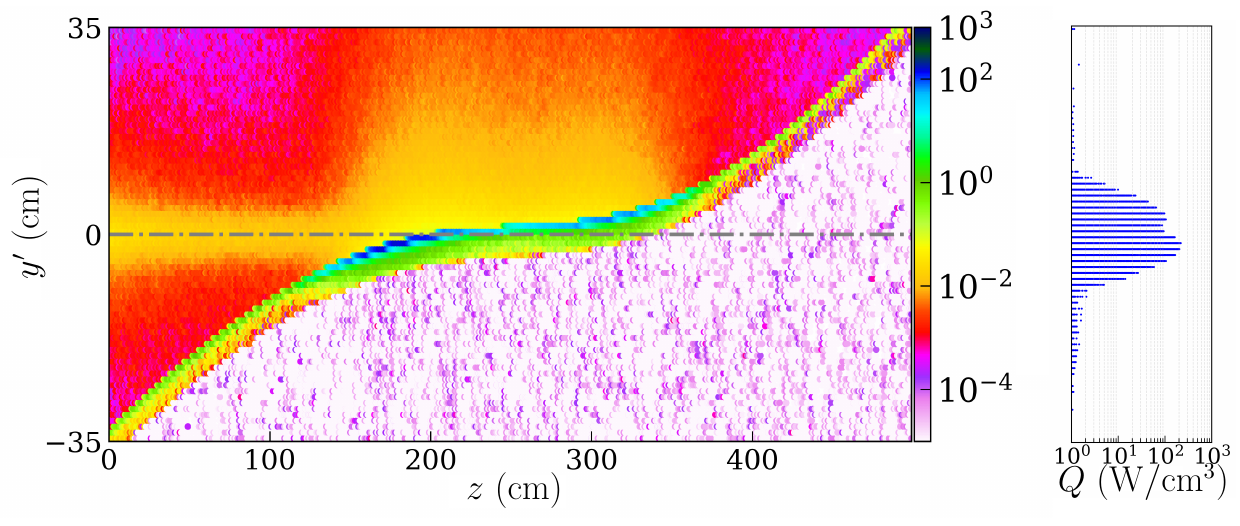}
         \caption{}
         \label{fig:5b}
     \end{subfigure}
     \hfill
        \caption{Power deposition of the photon beam on the (a)~L-slope and (b)~S-slope.}
        \label{fig:5}
\end{figure}

Monte Carlo simulations using the \textsc{fluka} code were performed to evaluate the power-density distribution for each slope design. The resulting power deposition at the midplane cross section is shown in \cref{fig:5}.

In the linear case [\cref{fig:5a}], the energy is concentrated within a narrow region, producing a peak power density of approximately \SI{500}{\watt\per\centi\meter\cubed}. This leads to a highly localized heating zone. In contrast, the S-slope [\cref{fig:5b}] spreads the energy over a broader area due to its curvature, resulting in a reduction in peak power density to around \SI{250}{\watt\per\centi\meter\cubed}. The broader spread in the curved shape is expected to reduce thermal gradients and improve absorber stability.

Quantitatively, the linear slope exhibits a single peak of approximately \SI{500}{\watt\per\centi\meter\cubed}, whereas the S-slope produces two smaller peaks of around \SI{250}{\watt\per\centi\meter\cubed} due to the angle of its interaction with the beam. This distribution improves heat dissipation by avoiding a single concentrated hot spot. The power absorbed by the lead is \SI{281.9}{\kilo\watt} for the linear slope and \SI{241.3}{\kilo\watt} for the S-slope, representing 76.2\% and 65.4\%, respectively, relative to the full \SI[number-unit-separator=\text{-}]{370}{\kilo\watt} beam power.

Despite the improved energy deposition achieved by the S-slope design, as can be seen in \cref{fig:5b}, some of the photon energy is reflected upward (backscattering), around \SI{63}{\kilo\watt} for Z-pole operation (17\% of the total power). This backscattering effect not only reduces the efficiency of energy absorption but also places additional demands on the upper region of the dump, which must be designed to handle this contribution.

\subsection{Thermo-fluid calculations for the slope concept}

Using the calculated power-deposition distributions, transient thermo-fluid simulations with conjugate heat transfer were conducted as outlined in Sec.~\ref{sec:3}.

Two characteristic timescales govern the system's thermal and flow evolution~\cite{LIU2023361, SANTOS2023168791}: the interfacial relaxation time and the viscous diffusion time, respectively defined as
\begin{equation}
\tau_\gamma = \sqrt{\frac{\rho L_{\mathrm{h}}^3}{\gamma}} \approx 1 \text{~s} \quad \mathrm{and} \quad \tau_\mu = \frac{L_{\mathrm{h}}^2}{\nu} \approx 1.6 \text{~h} .
\end{equation}
These timescales indicate how rapidly the interface responds to surface-tension disturbances ($\tau_{\gamma}$) and how momentum diffuses across the film ($\tau_{\mu}$). For transient simulations, the short interfacial time is the more relevant timescale.

The simulations are conducted as follows: (i)~$[-3~\mathrm{s},\,0~\mathrm{s}[$: transient flow phase, allowing the slope to be wetted by the liquid lead and reach a steady-state velocity; (ii)~$[0~\mathrm{s}]$: activation of the energy-source term to account for energy deposition; (iii)~$]0~\mathrm{s},\,10~\mathrm{s}[$: transient calculation phase, allowing the system to reach a steady-state temperature and velocity profile.

\begin{figure}[!htb]
     \centering
         \includegraphics[width=\textwidth]{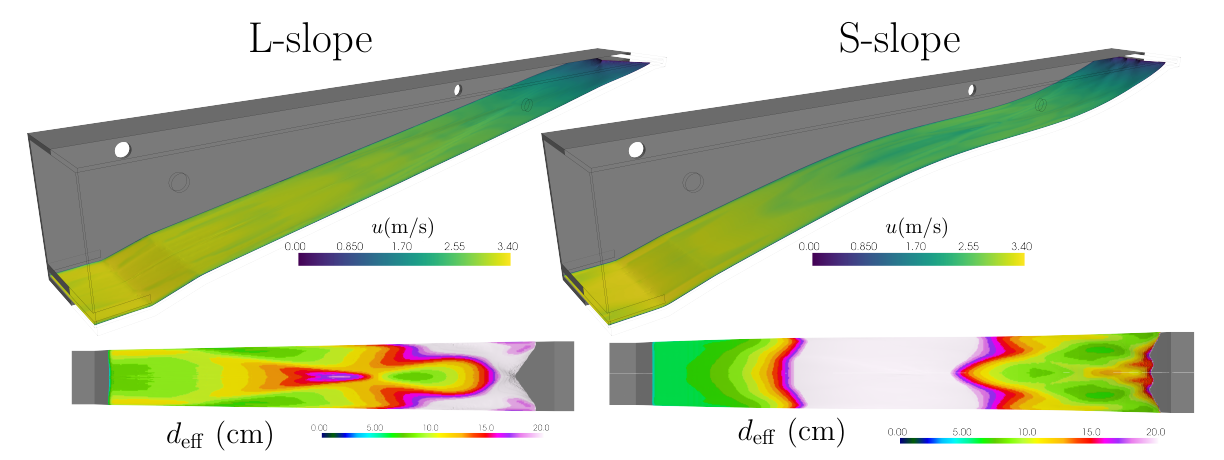}
         \caption{Velocity magnitude of the free surface of the liquid Pb (top) and its effective thickness (bottom) along the slope.}
         \label{fig:6a}

\end{figure}

\Cref{fig:6a} compares the velocity distribution of the free surface of the liquid lead and the corresponding effective thickness $d_{\text{eff}}$ for the L- and S-slope geometries.

In the L-slope configuration, the free-surface flow remains relatively uniform along the absorber, with a narrow region of increased velocity close to the inlet. This leads to localized thinning, as shown by the central low-$d_{\text{eff}}$ band. In contrast, the S-slope induces a redistribution of the flow that broadens the high-velocity region downstream. As a result, the effective thickness is more uniform and substantially greater across the photon footprint. From a thermal-hydraulic perspective, the more homogeneous velocity field of the S-slope also reduces the presence of stagnation zones, thereby supporting more uniform convective heat removal. Overall, these results demonstrate that the S-slope geometry provides superior performance in terms of fluid flow, absorber thickness, and photon-energy dissipation compared to the simpler L-slope design.

\begin{figure}[!htb]
     \centering
         \includegraphics[width=\textwidth]{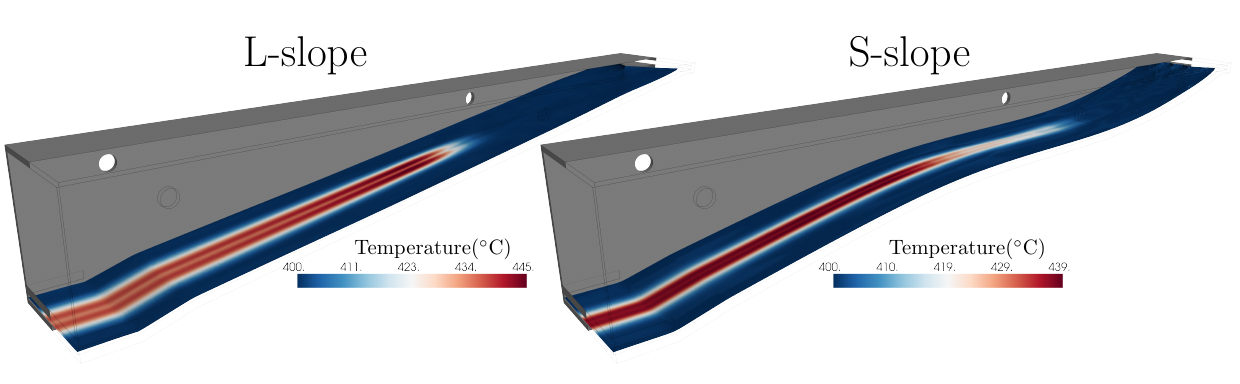}
         \caption{Steady-state temperature distribution at the liquid-Pb free surface for each of the slope profiles.}
        \label{fig:6b}
\end{figure}

\Cref{fig:6b} presents the temperature distribution on the free surface of the liquid lead for the L- and S-slope absorbers. In both cases, the bulk flow temperature remains close to the inlet condition of \SI{400}{\celsius}, with localized heating along the photon-impact line, leading to an increase of around 10--\SI{15}{\celsius} at the outlet.

The L-slope exhibits a more focused hot spot, reaching approximately \SI{445}{\celsius}, reflecting the region of lower effective thickness that can be observed in \cref{fig:6a}. In contrast, the S-slope distributes the deposited energy more evenly, with maximum surface temperatures remaining below \SI{440}{\celsius}. This confirms that the improved flow redistribution and higher $d_{\text{eff}}$ of the S-slope not only improve photon absorption but also mitigate localized thermal peaks. The corresponding reduction in temperature gradients is favorable for both thermal-hydraulic stability and the long-term compatibility of liquid lead with the structural walls, indicating that the S-slope is the more robust absorber configuration.

\begin{figure}[!htb]
    \centering
    \includegraphics[width=\linewidth]{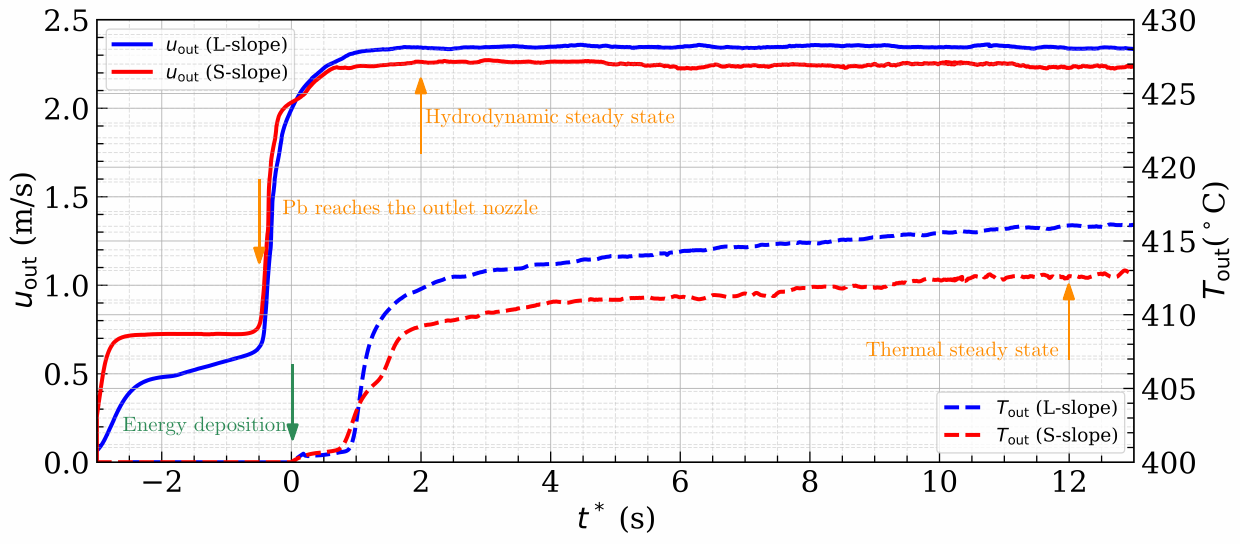}
    \caption{Temporal evolution of the velocity and mean temperature of the liquid-lead outlet, for the L- and S-slope cases.}
    \label{fig:8}

\end{figure}

\Cref{fig:8} shows the transient evolution of the flow velocity at the outlet boundary and the mean temperature of the liquid lead for the L- and S-slope cases. For numerical stability and accuracy, the transient simulations are performed in two distinct stages. First, an isothermal hydrodynamic calculation is carried out with the energy-source term disabled. During this phase, liquid lead is injected at the inlet and the solution is advanced until the flow fully covers the slope and reaches the outlet boundary ($t^*=-0.5$) with a steady hydrodynamic profile. Only after this condition is achieved is the volumetric power-deposition source (representing the BS energy input) activated ($t^*=0$).

At $t^{*}\approx -0.5$, both geometries show a brief increase in outlet velocity associated with the inflowing momentum of the liquid advancing down the slope. Shortly thereafter, the flow stabilizes at a hydrodynamic steady state of $u_{\text{out}}\approx \SI{2.2}{\meter\per\second}$ for both geometries.

\begin{figure}[!htb]
     \centering
     \begin{subfigure}[b]{0.49\linewidth}
         \centering
         \includegraphics[width=\textwidth]{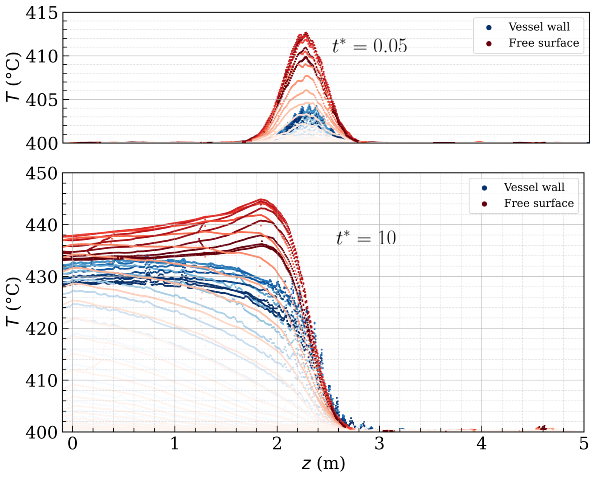}
         \caption{}
         \label{fig:7a}
     \end{subfigure}
     \hfill
     \begin{subfigure}[b]{0.49\linewidth}
         \centering
         \includegraphics[width=\textwidth]{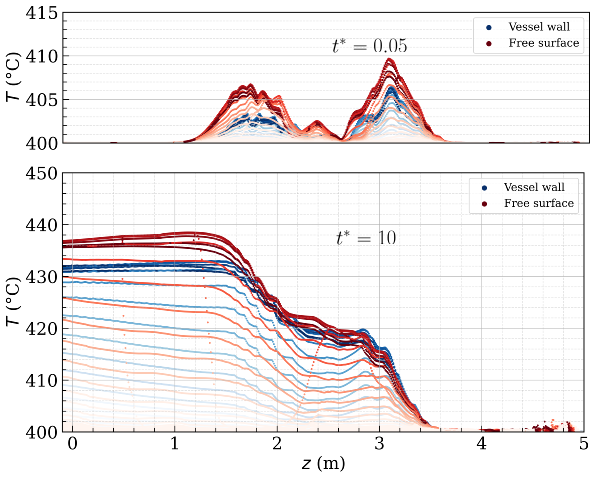}
         \caption{}
         \label{fig:7b}
     \end{subfigure}
     \hfill
        \caption{Temperature distribution of the free surface of the liquid lead and the vessel wall below for the (a)~L-slope and (b)~S-slope. In each case, the color intensity represents the distance from the origin, with full color at $x=\SI{0}{\meter}$ and transitioning to a lower-intensity color at $x=\SI{0.35}{\meter}$.}
        \label{fig:7}
\end{figure}

At $t^{*}=0$, BS power deposition is initiated. The outlet temperature responds slowly, reflecting the large thermal inertia of the liquid-lead volume. The L-slope slope eventually settles at a slightly higher steady-state outlet temperature (approximately \SI{416}{\celsius}) than the S-slope (approximately \SI{413}{\celsius}). This trend is consistent with the more uniform energy distribution and reduced hot-spot formation in the S-slope configuration (see \cref{fig:7}). The separation between the fast hydrodynamic convergence and the slower thermal equilibration highlights an important operational feature: stable flow conditions are established well before the system reaches a steady thermal state, which increases robustness during beam transients.

\Cref{fig:7} illustrates the temperature distribution on the liquid-lead surface (red color map) and the steel wall (blue color map) shortly after power deposition begins ($t^{*}=0.05$) and at thermal steady state ($t^{*}=10$). At early times, the L-slope exhibits a pronounced localized hot spot, while the S-slope shows a more distributed heating pattern due to its curved geometry intersecting the photon path over a greater area. As expected, the peak temperature in the S-slope configuration remains lower. Both geometries stay below \SI{450}{\celsius} once steady-state conditions have been reached, comfortably satisfying the operational temperature limit of \SI{500}{\celsius} near the wall. In absolute terms, the L- and S-slope configurations both remain around \SI{80}{\celsius} below the temperature threshold of the design.

To verify the applicability of the turbulence model, the nondimensional wall distance, $y^+$, was monitored. This quantity is commonly used in CFD to assess whether the grid point closest to the wall lies in the appropriate region of the boundary layer for the selected turbulence model, ensuring that wall functions receive physically consistent input and that near-wall shear and heat transfer are modeled reliably. With a wall-function approach suitable for weak turbulence, the simulation maintained $y^+$ values between 30 and 100, which is appropriate for accurate near-wall treatment (see Appendix~\ref{app:B}).

\subsection{Double-slope concept for backscattering mitigation}

\begin{figure}[!htb]
    \centering
    \includegraphics[width=0.5\linewidth]{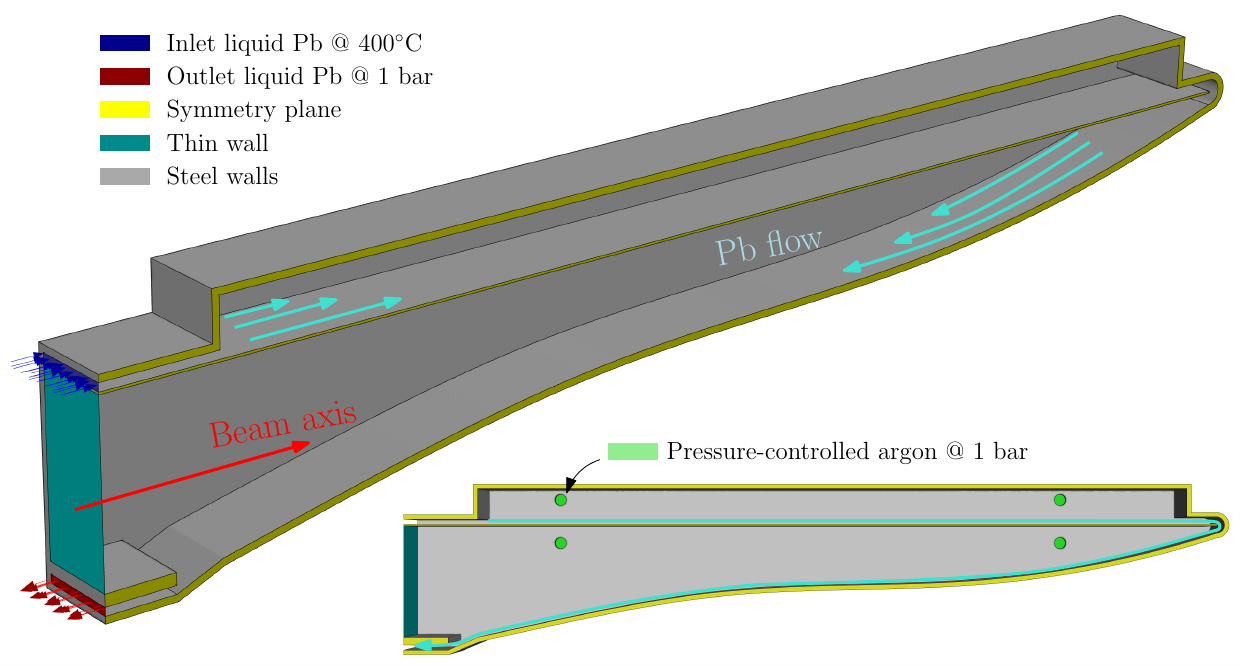}
    \includegraphics[width=0.4\linewidth]{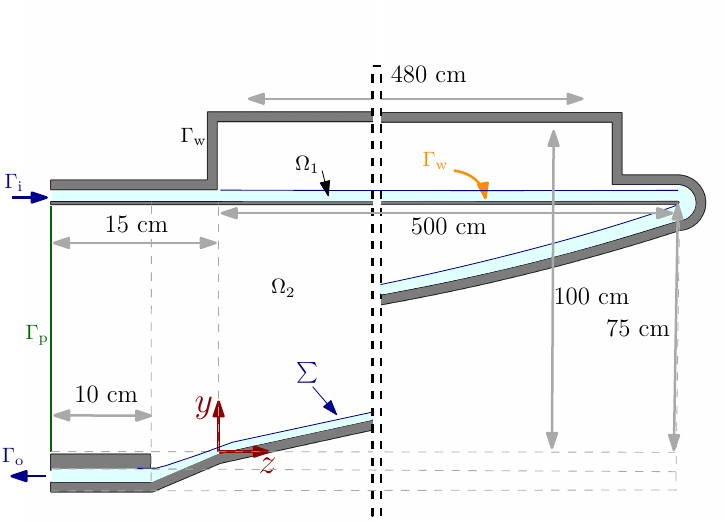}
    \caption{Schematic showing the double-slope design proposed to mitigate the backscattering problem (left) and a diagram presenting its parameters and boundary conditions (right). The symbols $\Gamma$ represent the boundary conditions for the inlet (i), outlet (o), and walls (w), and $\sum$ indicates the interface (buffer zone) between the liquid phase ($\Omega_1$) and the gas phase ($\Omega_2$).}
    \label{fig:9}

\end{figure}

To address the issue of upward-backscattered radiation (\cref{fig:5}), a double-slope design (see \cref{fig:9}) was proposed. In this configuration, the liquid lead first flows over an upper slope before passing over an S-slope design (as previously described). An upper steel plate is added at a height of $H_{\text{d}}=$ \SI{75}{\centi\meter}, along with an upper cover wall at $H_{\text{c}}=$ \SI{100}{\centi\meter}.

The purpose of this design is twofold: as established above, the use of the S-slope increases the interaction depth for initial photon absorption; however, the inclusion of the upper slope now serves to absorb any backscattered photons while maintaining the temperature of the upper steel within a stable range. The upper steel plate and the layer of liquid lead on the upper slope effectively absorb the reflected energy, preventing it from escaping the dump. The upper steel plate is taken to have a \SI[number-unit-separator=\text{-}]{1}{\centi\meter} thickness, thin enough to avoid significant interaction with the photon radiation but thick enough to structurally support the flow of liquid lead above it. This upper plate will then support a volume of liquid lead of roughly $5 \times \SI{0.7}{\meter\squared}$, with a thickness of approximately 2--\SI{3}{\centi\meter}, corresponding to about \SI{1}{\tonne} of weight.

\begin{figure}[!htb]
    \centering
    \includegraphics[width=\linewidth]{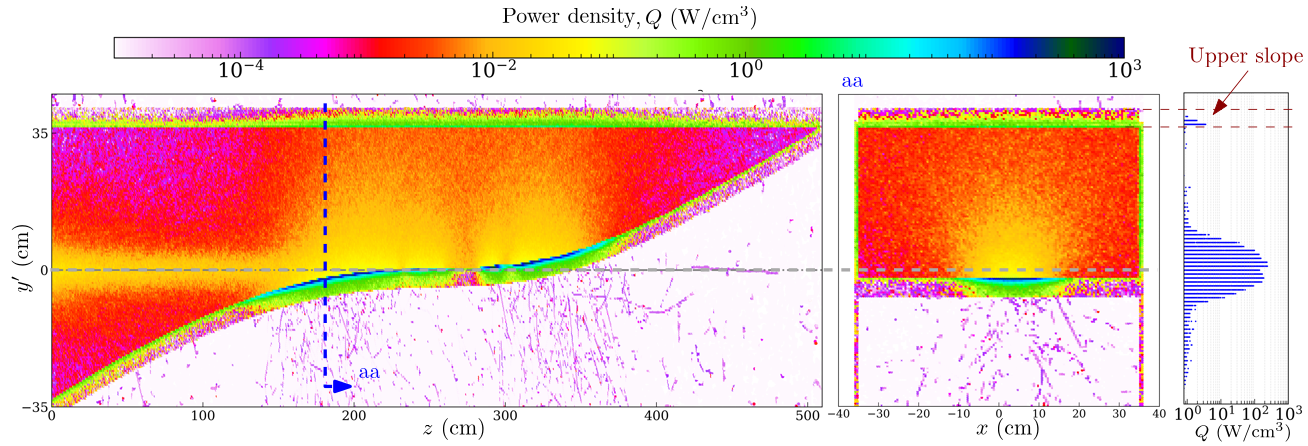}
    \caption{Power-deposition maps for the photon beam in the double-slope design, showing $zy$ (left) and $xy$ (right) cross-section planes.}
    \label{fig:11}

\end{figure}

Having defined this concept geometrically, a \textsc{fluka} Monte Carlo simulation was performed to obtain the power deposition for the multiple components of the configuration. The results for the most energetic cut plane (the center of the beam spot) can be seen in \cref{fig:11}.

\begin{figure*}[!htb]
    \centering
    \includegraphics[width=0.9\linewidth]{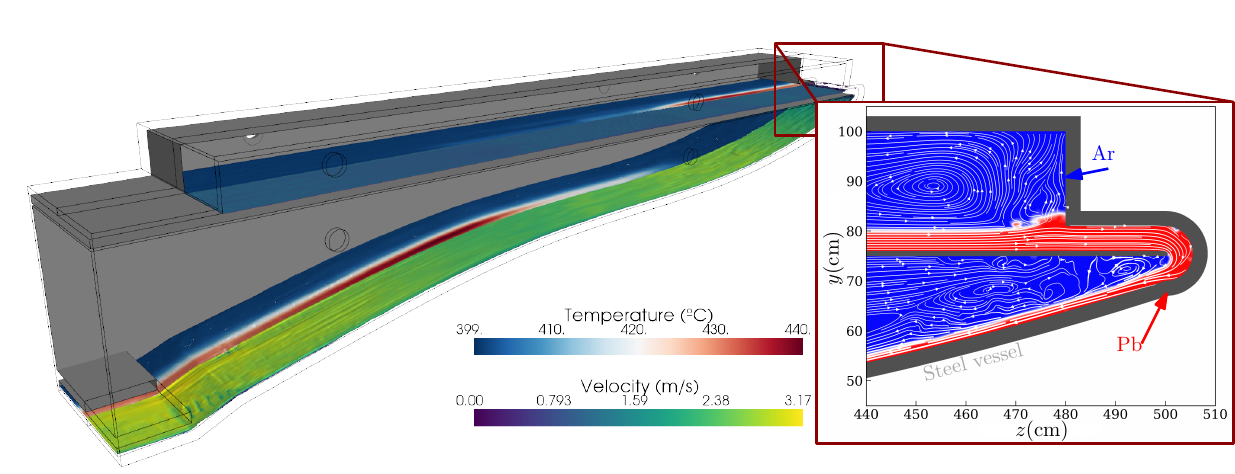}
    \caption{Three-dimensional visualization of the temperature field and instantaneous velocity field of the free surface of the liquid Pb, showing an enlarged view of the instantaneous stream lines in the elbow region.}
    \label{fig:12}
\end{figure*}

From a fluid-dynamic point of view, it was intended that the double-slope design should produce a smooth flow across the two slopes. Therefore, a channel and tangential elbow are included at the end of the top slope to redirect the flow without generating significant additional turbulence losses or significantly constraining the flow. It can be seen in \cref{fig:12} that the upper flow maintains a low-velocity profile and the lower slope has the same smooth interface transition as the previous single slope.

In this design, the upper slope is capable of absorbing all the backscattered radiation: as can be seen in \cref{fig:11}, the power density is zero above the upper flow of lead. In Z-pole operation, this results in the upper steel plate absorbing around \SI{36.45}{\kilo\watt}. The temperature of this steel plate is then maintained in a stable equilibrium due to the flow of lead along the upper slope. \Cref{fig:12} clearly illustrates the heating of the upper flow, and this heat is then displaced with the velocity field of the Pb.


In this design, as shown in the inset of \cref{fig:12}, the liquid-lead flow exhibits a recirculation zone at the entrance to the elbow, which can be described as a region where the flow accumulates and forms vortices due to the sudden change in direction. This phenomenon is a result of adverse pressure gradients and the sharp curvature of the elbow. The fluid inertia, combined with centrifugal forces, leads to flow separation, creating a low-velocity recirculation region or ``stagnation region,'' where the lead accumulates momentarily before rejoining the main flow downstream.

\begin{figure}[!htb]
     \centering
     \begin{subfigure}[b]{0.49\linewidth}
         \centering
          \includegraphics[width=\linewidth]{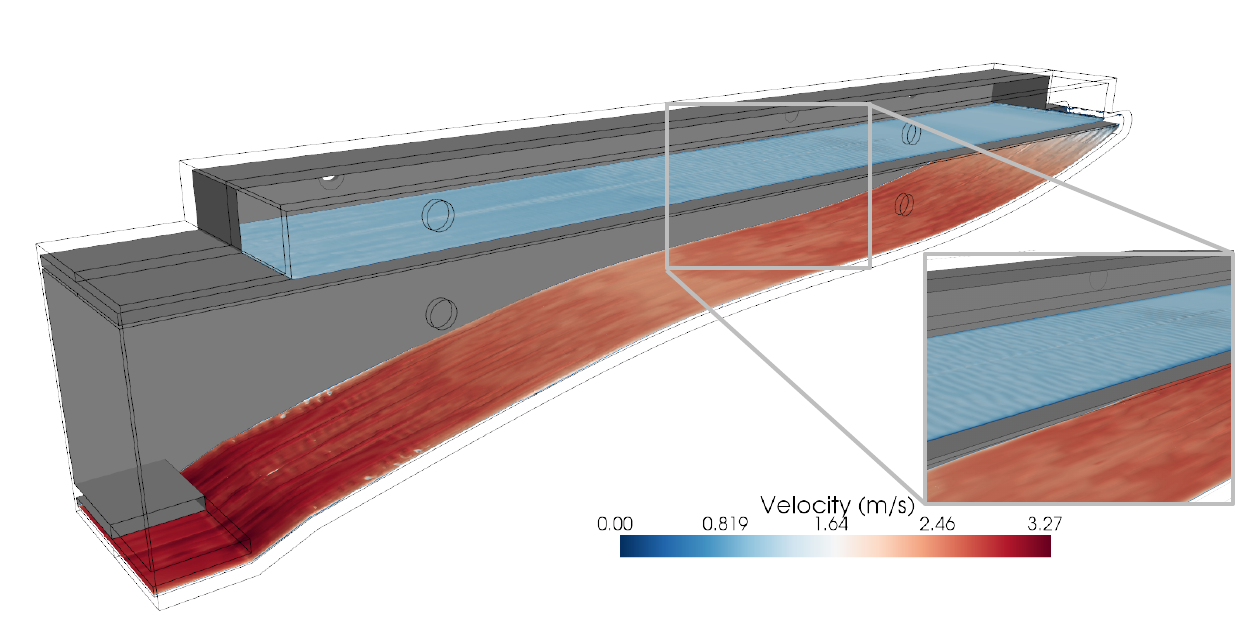}
         \caption{}
         \label{fig:13a}
     \end{subfigure}
     \hfill
     \begin{subfigure}[b]{0.49\linewidth}
         \centering
         \includegraphics[width=\linewidth]{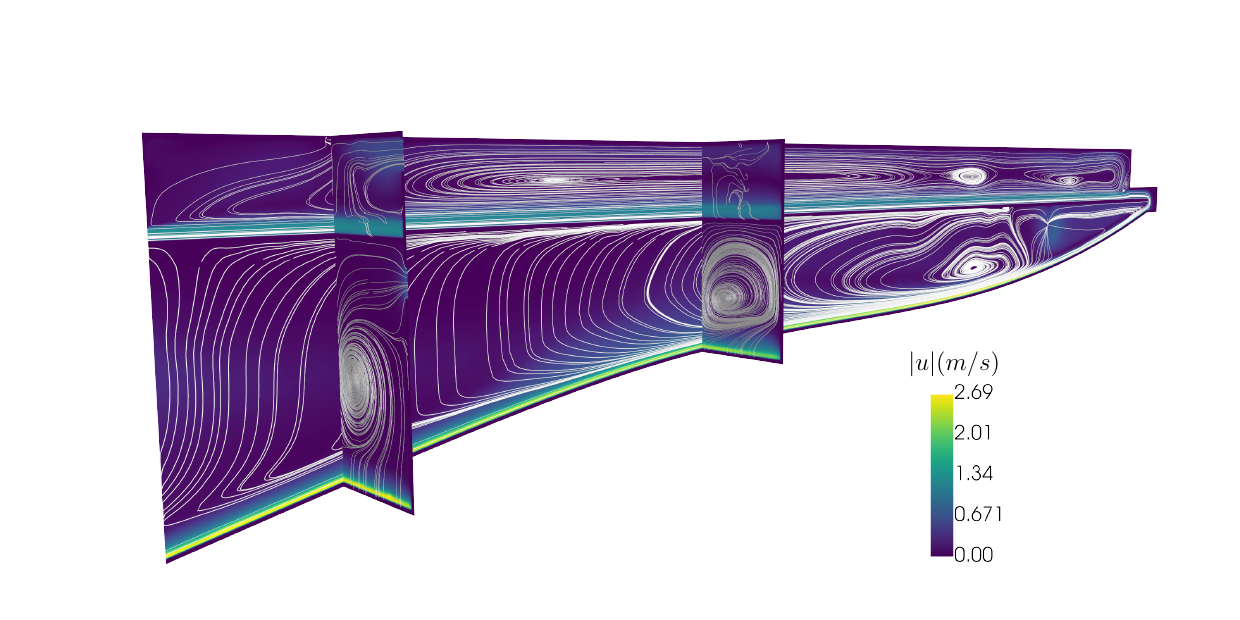}
         \caption{}
         \label{fig:13b}
     \end{subfigure}
     \hfill
        \caption{(a)~Velocity field of the free surface of the liquid Pb and (b)~velocity-magnitude distributions on three different cutting planes, showing the streamlines of the vector field in each case.}
        \label{fig:13}
\end{figure}

This flow behavior is particularly critical in the elbow, where the transition from a linear to a curved path increases the likelihood of flow separation, causing a localized buildup of liquid lead and generating turbulence. This effect is clearly visible in \cref{fig:13}, where it can be seen that the recirculation region disrupts the smooth flow profile, which could contribute to increased thermal stresses or inefficiencies.

One of the critical engineering challenges of the liquid-lead dump is thermal management. The absorption of high-energy BS photons will generate significant heat within the dump structure. To ensure that the steel components do not exceed safe temperature limits, we need to evaluate the thermal behavior of the critical steel plates. Based on the results of this analysis, it is then essential to assess the thermal feasibility of the design, focusing primarily on two aspects: the temperature rise in the liquid lead and the power absorbed by the steel walls, particularly the upper slope, which is subjected to continuous direct impact from backscattered radiation and is cooled by the upper flow of liquid lead.

\begin{figure}[!htb]
     \centering
     \begin{subfigure}[b]{0.49\linewidth}
         \centering
         \includegraphics[width=\textwidth]{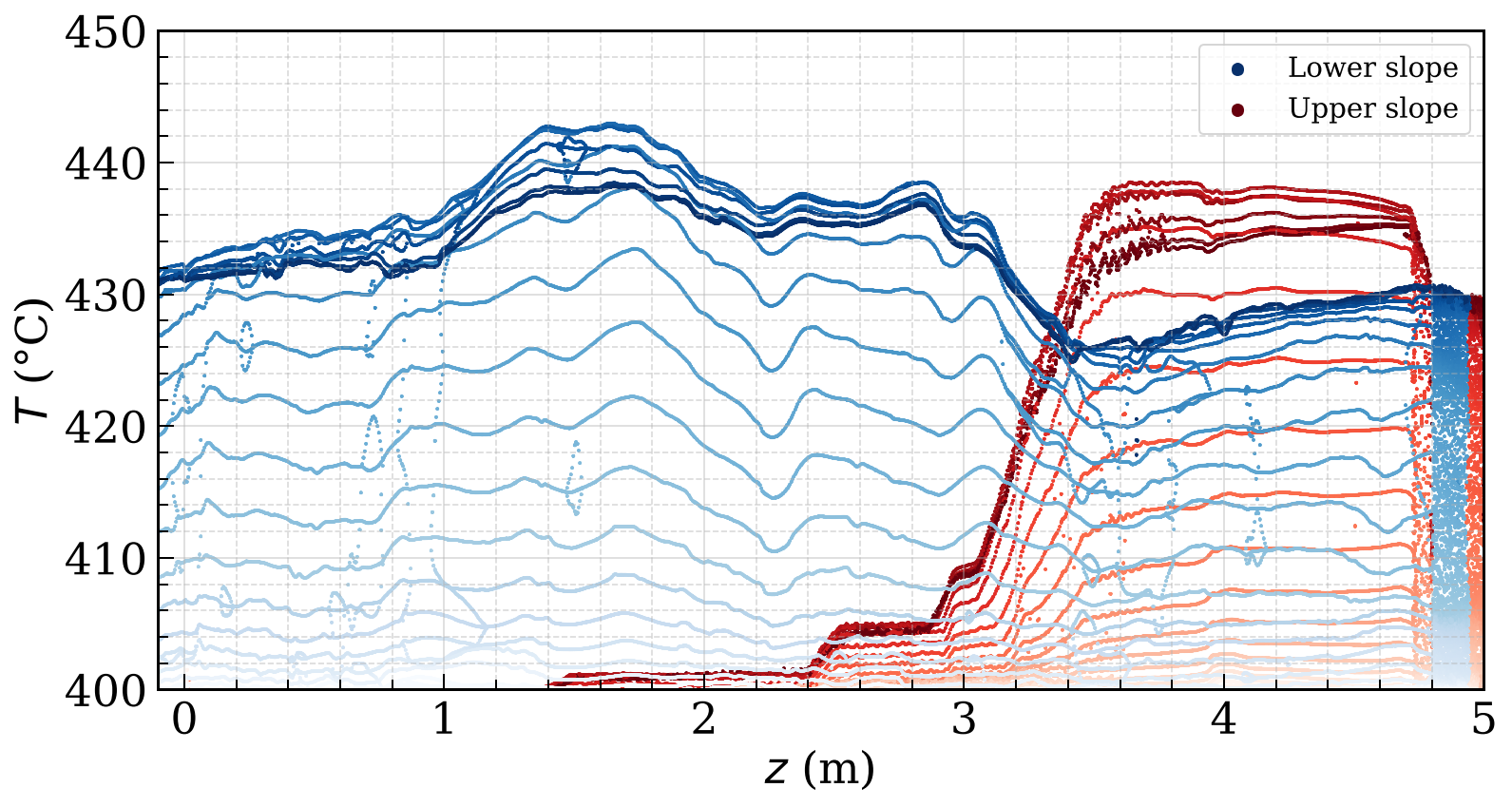}
         \caption{}
         \label{fig:14a}
     \end{subfigure}
     \hfill
     \begin{subfigure}[b]{0.49\linewidth}
         \centering
         \includegraphics[width=\textwidth]{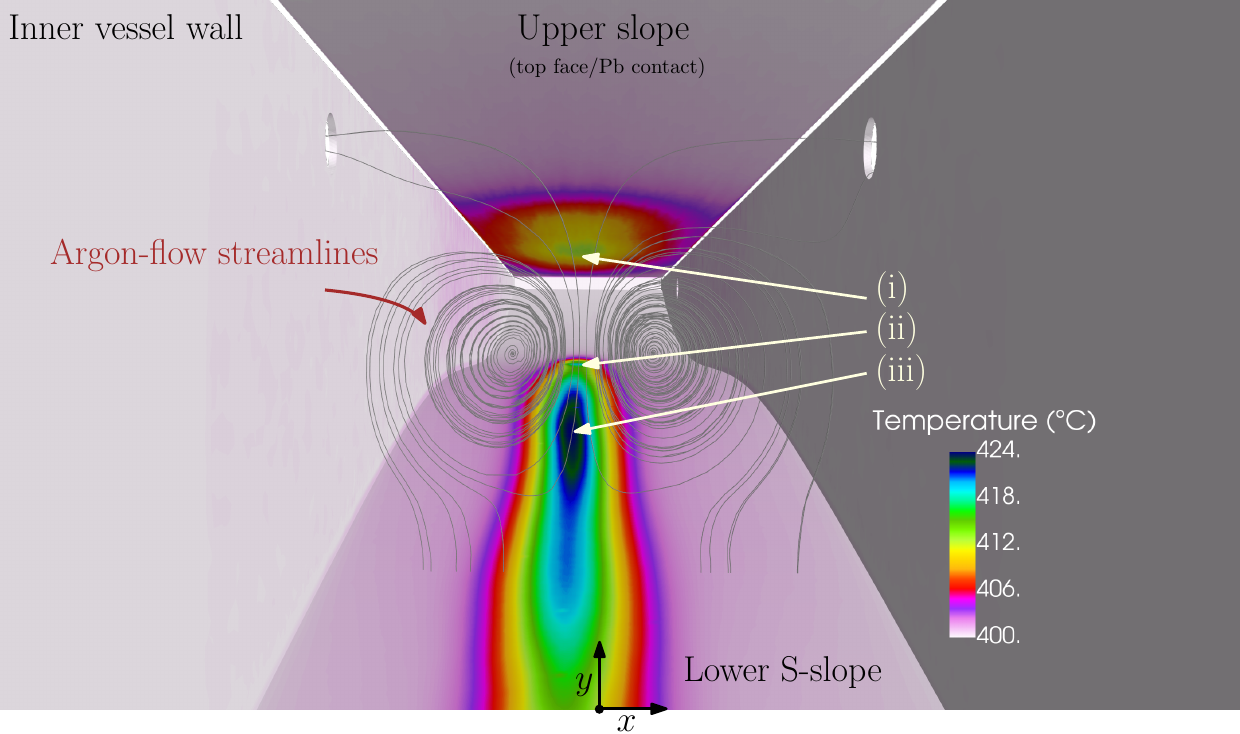}
         \caption{}
         \label{fig:14b}
     \end{subfigure}
     \hfill
        \caption{(a)~Plots showing the temperature distribution along the free surface of the lead in the upper region, in red, and lower region, in blue. As in \cref{fig:7}, the color intensity represents the distance from the origin, with full color at $x=\SI{0}{\meter}$ and transitioning to a lower-intensity color at $x=\SI{0.35}{\meter}$. (b)~Three-dimensional view of the double-slope configuration, looking along the beam axis, with superimposed color maps showing the temperature distribution on the steel vessel. Points~(i), (ii), and (iii) indicate the main hot spots created at the vessel wall. }
        \label{fig:14}
\end{figure}

\Cref{fig:14b} presents a three-dimensional view of the liquid-lead dump, looking along the beam axis, with superimposed color maps showing the temperature distribution on the steel vessel. We can observe that the main increase in temperature occurs in the S-slope, as expected, and the temperature of the upper slope increases by around \SI{10}{\celsius}. There is also some degree of temperature increase on the side walls, which absorb around \SI{6.8}{\kilo\watt} (1.8\%) of the total energy. Furthermore, it is clear that due to the transverse offset, the left-hand side wall [of the view in \cref{fig:14b}] exhibits a slightly higher increase in temperature, a difference of around \SI{2}{\celsius}.

The results of these simulations show that the liquid lead provides effective cooling to the upper steel plate. The upper layer of liquid lead, particularly in the S-slope region, serves not only to absorb backscattered photons but also to dissipate the heat generated by the absorbed energy. This cooling mechanism prevents thermal damage to the steel, ensuring the structural integrity of the dump over extended operation periods.

Nonetheless, this inclined-slope concept also has some drawbacks. Due to its large footprint and high mass, integrating it into the tunnel may present considerable difficulties, and its large free surface also increases the risk of contamination and liquid-metal oxidation. These factors motivated the development of a more compact design with the goals of optimizing overall system operation and improving ease of integration.

\subsection{Compact upstream slope plus pool (CUSP)}

The double-slope configuration effectively mitigates backscattering, but it is mechanically bulky and complex to integrate into the tunnel. In addition, its large free surface increases the risks of oxidation and contamination. To address these issues, a more compact alternative, referred to as the ``compact upstream slope plus pool'' (CUSP) configuration, was developed. This design retains the absorption efficiency of the sloped concept while drastically reducing its footprint and hydraulic complexity.

\begin{figure}[!htb]
    \centering
    \begin{subfigure}[b]{0.48\linewidth}
        \centering
        \includegraphics[width=\textwidth]{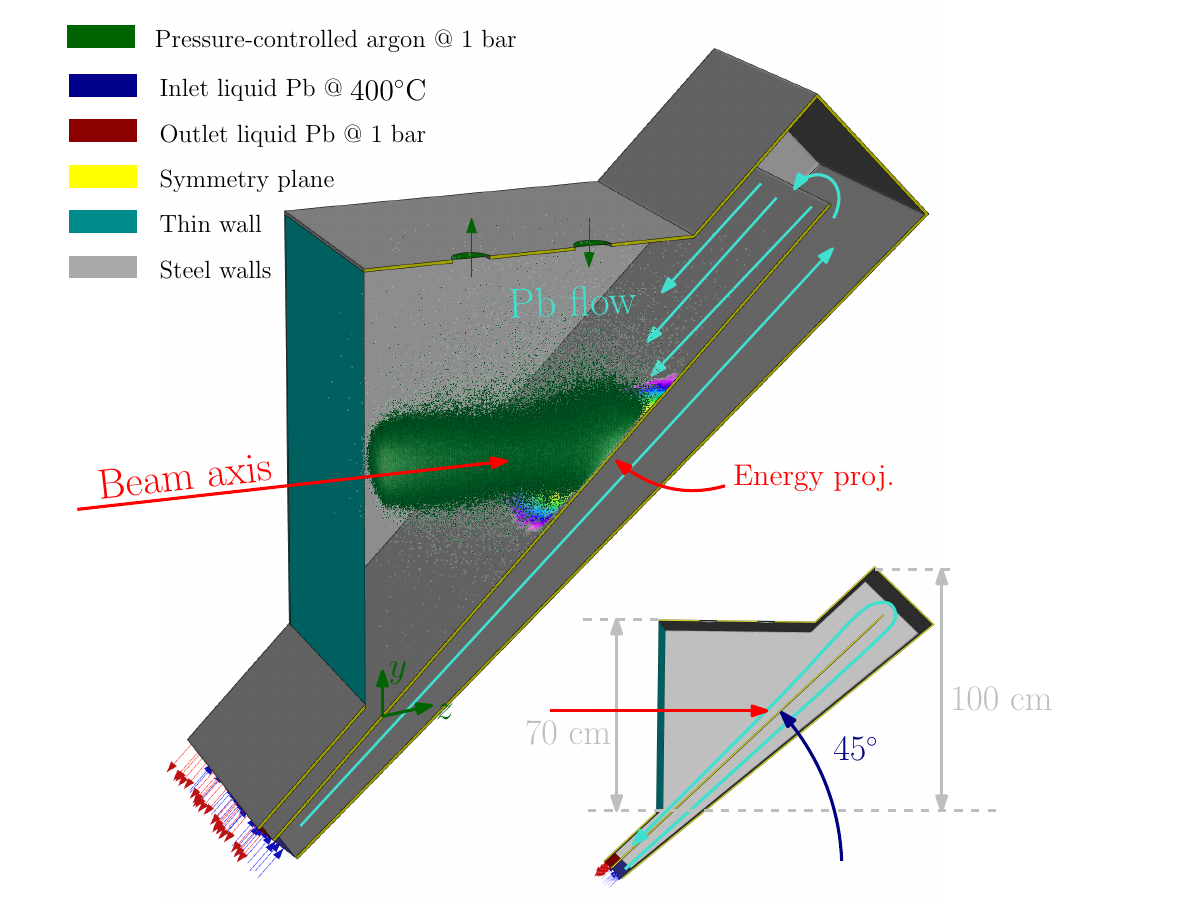}
        \caption{}
        \label{fig:15a}
    \end{subfigure}
    \begin{subfigure}[b]{0.5\linewidth}
        \centering
        \includegraphics[width=\textwidth]{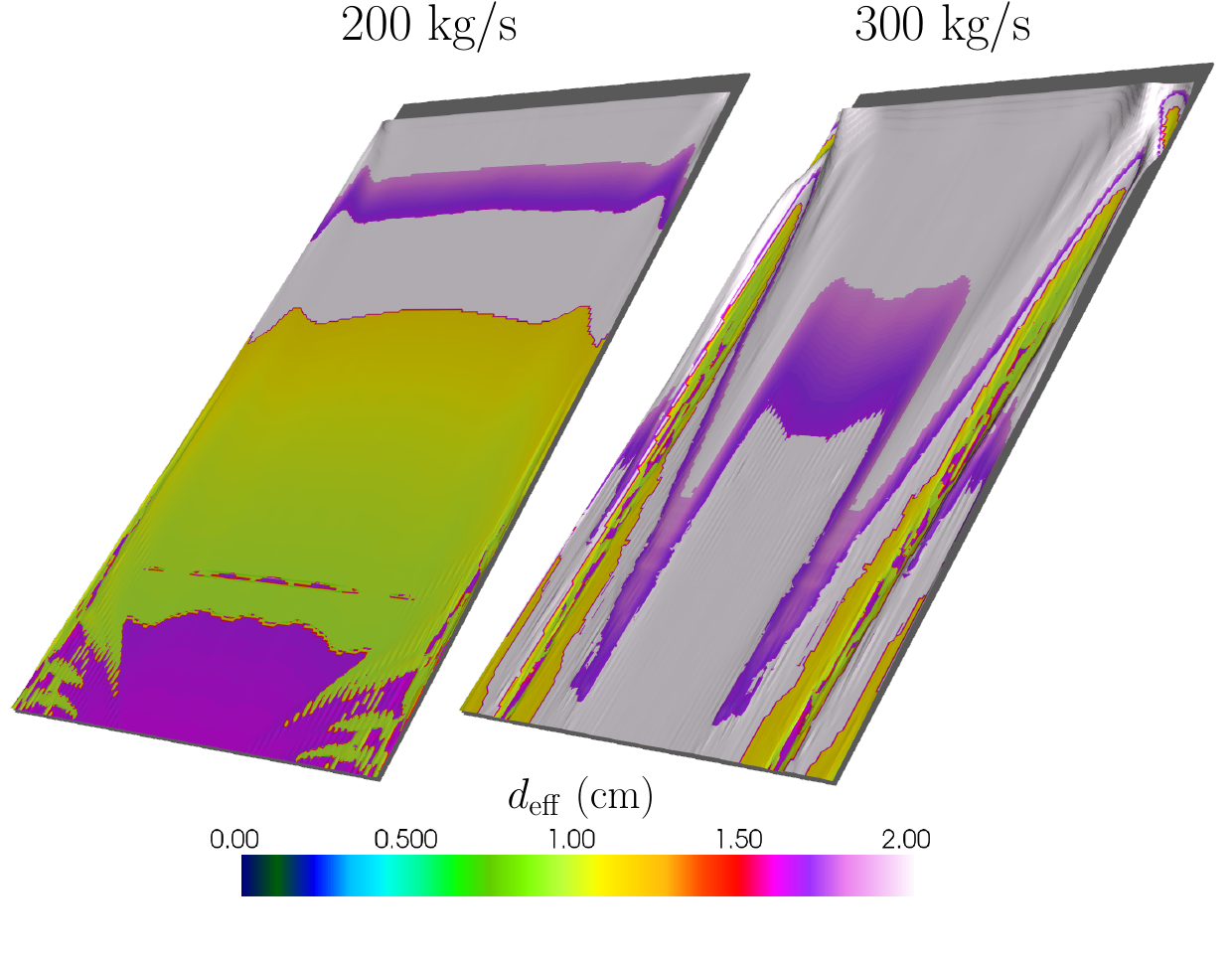}
        \caption{}
        \label{fig:15b}
    \end{subfigure}
    \caption{(a)~Schematic of the CUSP concept and (b)~three-dimensional plots showing the effective thickness for different liquid-lead mass flow rates.}
    \label{fig:15}

\end{figure}

The concept divides the absorption into two main regions: a thin, free-surface liquid-lead film that initially intercepts the photon beam's high-power peak, and a downstream quiescent lead pool to absorb the residual radiation. This configuration, shown schematically in \cref{fig:15a}, allows the most intense energy deposition to occur in a continuously renewed free-surface flow region, while the pool acts as a large thermal buffer.

In this configuration, the effective thickness and stability of the free-surface flow depend directly on the mass flow rate, as shown in \cref{fig:15b}. For a flow rate of \SI{300}{\kilogram\per\second}, the curtain achieves an effective thickness of approximately \SI{2}{\centi\meter}, whereas at \SI{200}{\kilogram\per\second} it is reduced to about \SI{1}{\centi\meter}.

\begin{figure}[!htb]
    \centering
    \begin{subfigure}[b]{0.49\linewidth}
        \centering
        \includegraphics[width=\textwidth]{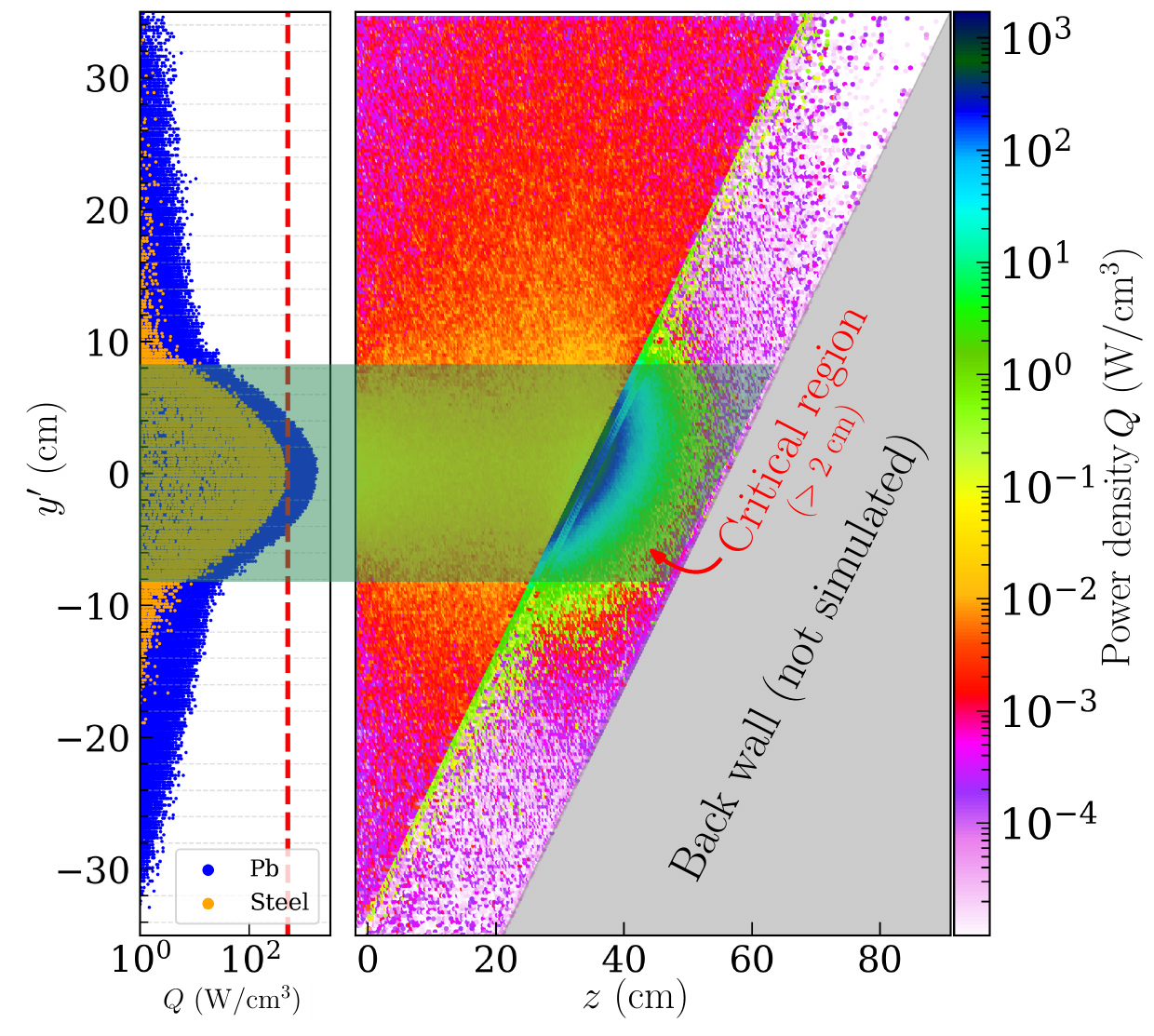}
        \caption{}
        \label{fig:16a}
    \end{subfigure}
    \begin{subfigure}[b]{0.49\linewidth}
        \centering
        \includegraphics[width=\textwidth]{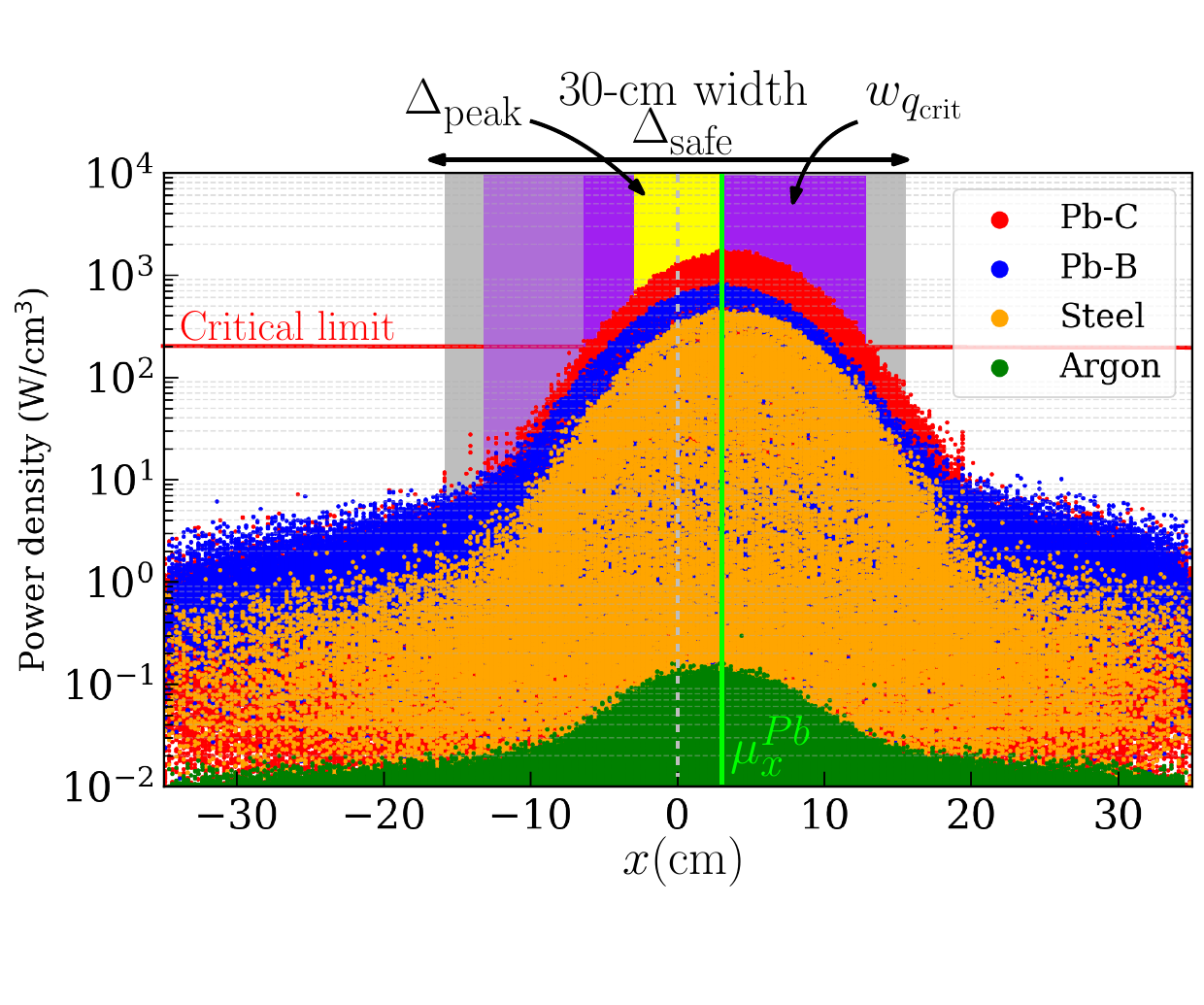}
        \caption{}
        \label{fig:16b}
    \end{subfigure}
    \caption{Power-density distribution in (a)~the $yz$ cut plane and (b)~the transverse plane. In~(b), $\Delta_{\mathrm{peak}}$ denotes the expected BS-peak displacement, $w_{q_\mathrm{crit}}$ is the width leading to steel temperatures above \SI{500}{\celsius}, and $\Delta_{\mathrm{safe}} \approx \SI{30}{\centi\meter}$ is the required upstream free-surface coverage.}
    \label{fig:16}
\end{figure}

\textsc{fluka} simulations (\cref{fig:16}) showed that a \SI[number-unit-separator=\text{-}]{2}{\centi\meter}-thick curtain is sufficient to absorb the peak photon power while keeping the volumetric power density in the adjacent steel below \SI{500}{\watt\per\centi\meter\cubed}. The remaining energy is deposited at lower intensity in the downstream pool, where it is fully absorbed without penetration. Furthermore, with a $45^\circ$ inclination, the results also indicate negligible backscattering within statistical uncertainty, eliminating the need for the upper absorption layer required by the double-slope design.

In this configuration, the upstream film absorbs about 53\% of the total photon power (roughly \SI{190}{\kilo\watt}), while the downstream pool absorbs around \SI{150}{\kilo\watt} and the steel vessel absorbs about \SI{16}{\kilo\watt}, giving a total absorption of \SI{356}{\kilo\watt}.

As previously noted, beam-beam effects may shift the photon footprint, creating a small transverse misalignment between the BS fan and the dump. To account for this, the main deposited power is conservatively assumed to lie within a \SI[number-unit-separator=\text{-}]{30}{\centi\meter}-wide region centered on the dump axis, corresponding to about 64\% of the total power [see \cref{fig:16b}]. In \cref{fig:16b}, each point represents a \textsc{fluka} scoring cell in the transverse plane at the curtain location, plotted as the local power density versus the horizontal coordinate~$x$.

The parameter $\Delta_{\mathrm{peak}}$ denotes the expected horizontal excursion of the BS peak due to optics errors and beam-beam offsets. Because the BS distribution is symmetric for each IP, this shift appears as $\pm \Delta_{\mathrm{peak}}$ around the nominal axis. The quantity $w_{q_\mathrm{crit}}$ identifies the width over which the local power density in the steel would cause the wall temperature to exceed \SI{500}{\celsius}. Based on this criterion, and including a safety margin, a central region of approximately $\Delta_{\mathrm{safe}} \approx \SI{30}{\centi\meter}$ is defined: if the upstream free-surface curtain fully covers this width, the peak power density in the steel will remain below the critical limit, keeping the vessel temperature safely below \SI{500}{\celsius}. From \cref{fig:16b}, it can be seen that the power density within this region stays below the critical value of \SI{250}{\watt\per\centi\meter\cubed}, even when accounting for a potential beam offset $\Delta_{\mathrm{peak}}$. This threshold corresponds to the maximum allowable deposition that maintains the steel wall below \SI{500}{\celsius}.

Based on these considerations, the free-surface flow was further optimized by narrowing its width, thus increasing its effective thickness via mass conservation. A key advantage of the CUSP geometry is that this thickness is achieved by geometrical design rather than by increasing the total flow rate. The concept can therefore operate safely within the nominal hydraulic envelope ($\dot{m} \lesssim \SI{300}{\kilogram\per\second}$).

\begin{figure}[!htb]
    \centering
    \begin{subfigure}[b]{0.49\linewidth}
        \includegraphics[width=\linewidth]{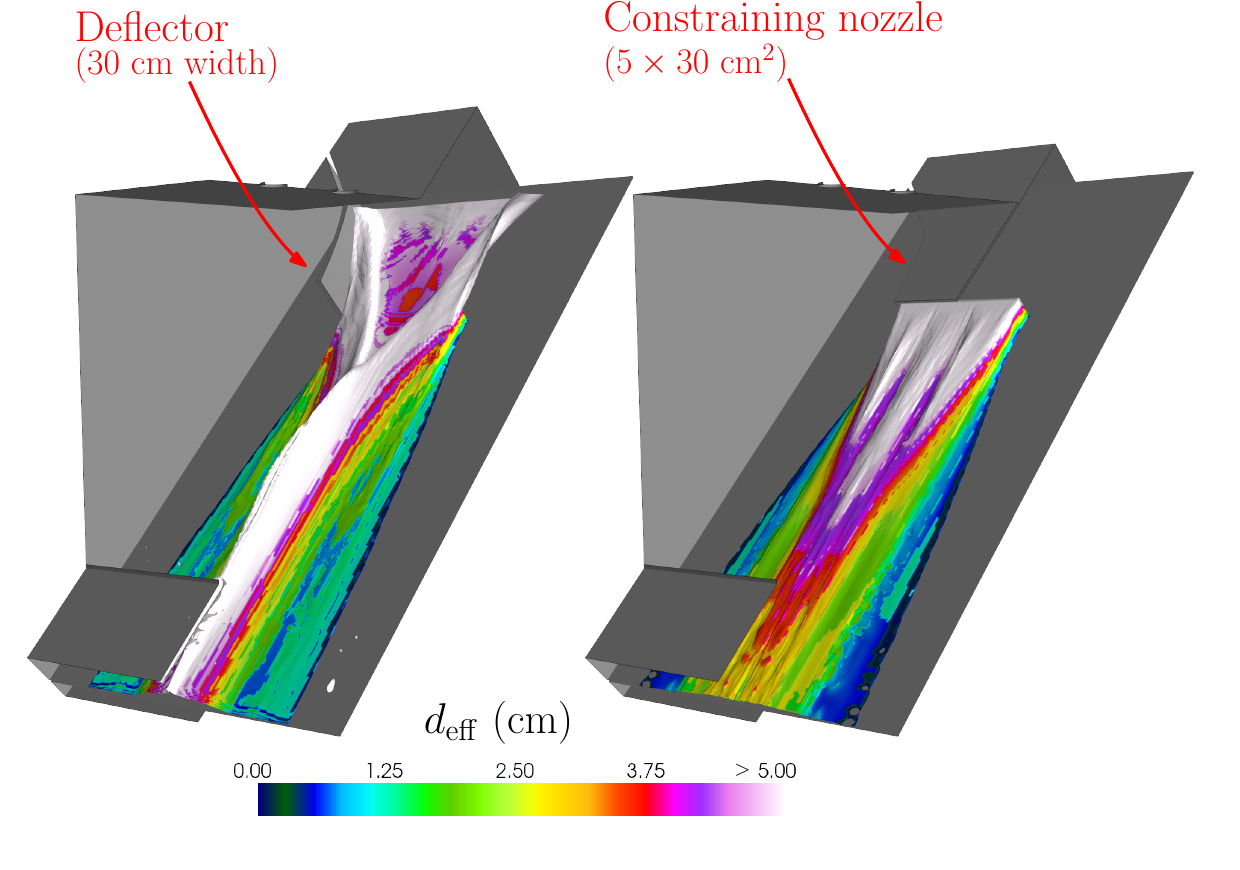}
        \caption{}
        \label{fig:18a}
    \end{subfigure}
    \begin{subfigure}[b]{0.49\linewidth}
        \includegraphics[width=\linewidth]{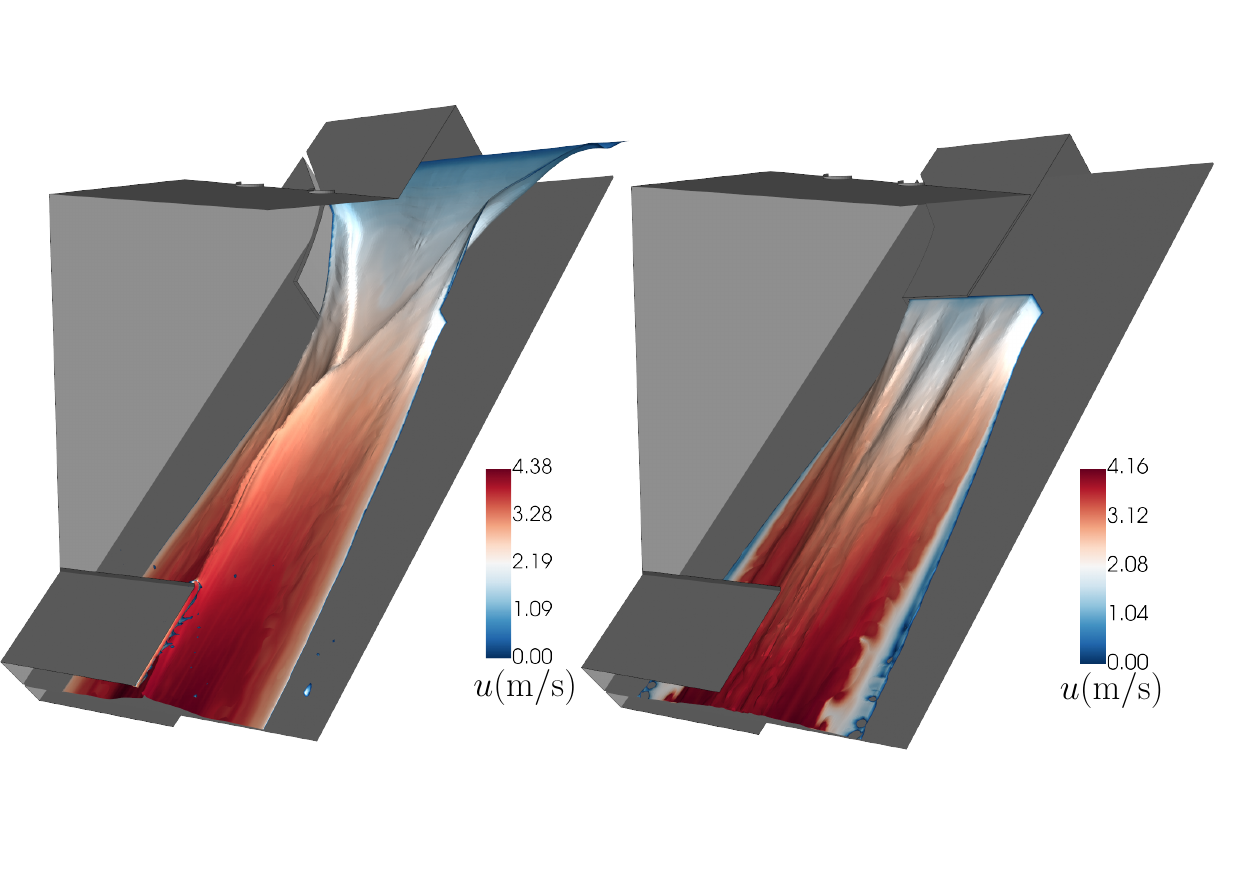}
        \caption{}
        \label{fig:18b}
    \end{subfigure}
    \caption{(a)~Effective thickness and (b)~velocity magnitude for constrained flow with either a \SI[number-unit-separator=\text{-}]{30}{\centi\meter}-wide deflector (left-hand image in each panel) or a $5\times\SI[number-unit-separator=\text{-}]{30}{\centi\meter\squared}$ constraining nozzle (right-hand image in each panel).}
    \label{fig:18}
\end{figure}

Two variants were investigated (\cref{fig:18}): (i)~a simple compact slope (as described above) and (ii)~a compact slope with either a \SI[number-unit-separator=\text{-}]{30}{\centi\meter}-wide deflector or a $5\times\SI[number-unit-separator=\text{-}]{30}{\centi\meter\squared}$ constraining nozzle.

The first constraint was introduced to reduce the transverse width of the free-surface overflow without adding a vertical front wall. This lateral redirection of the flow, however, induces wave-impact instabilities, as visible in the isosurfaces of \cref{fig:18a}. Nevertheless, the power-deposition results in \cref{fig:16a} show that the critical region is not strictly defined in the $x$ direction but also includes a narrow band in $y$, with the dominant contribution confined to $y\lesssim \SI{10}{\centi\meter}$. This implies that the free surface does not need to remain unconstrained across the full height of the vessel.

Based on this observation, the vessel wall was extended and a geometric constraint was added to limit the outgoing component of the velocity. This yields a nozzle that releases the free-surface liquid lead approximately \SI{12}{\centi\meter} from the peak-power location, providing a safe margin for expected beam-spot variations while significantly increasing the effective thickness of the absorber. As shown in \cref{fig:18}, the constraining nozzle ($5\times\SI{30}{\centi\meter\squared}$) produces a thicker, more stable film, with $d_\mathrm{eff}$ exceeding \SI{5}{\centi\meter} across most of the main footprint of the beam.

\begin{figure}[!htb]
    \centering
    \begin{subfigure}[b]{0.49\linewidth}
        \includegraphics[width=\linewidth]{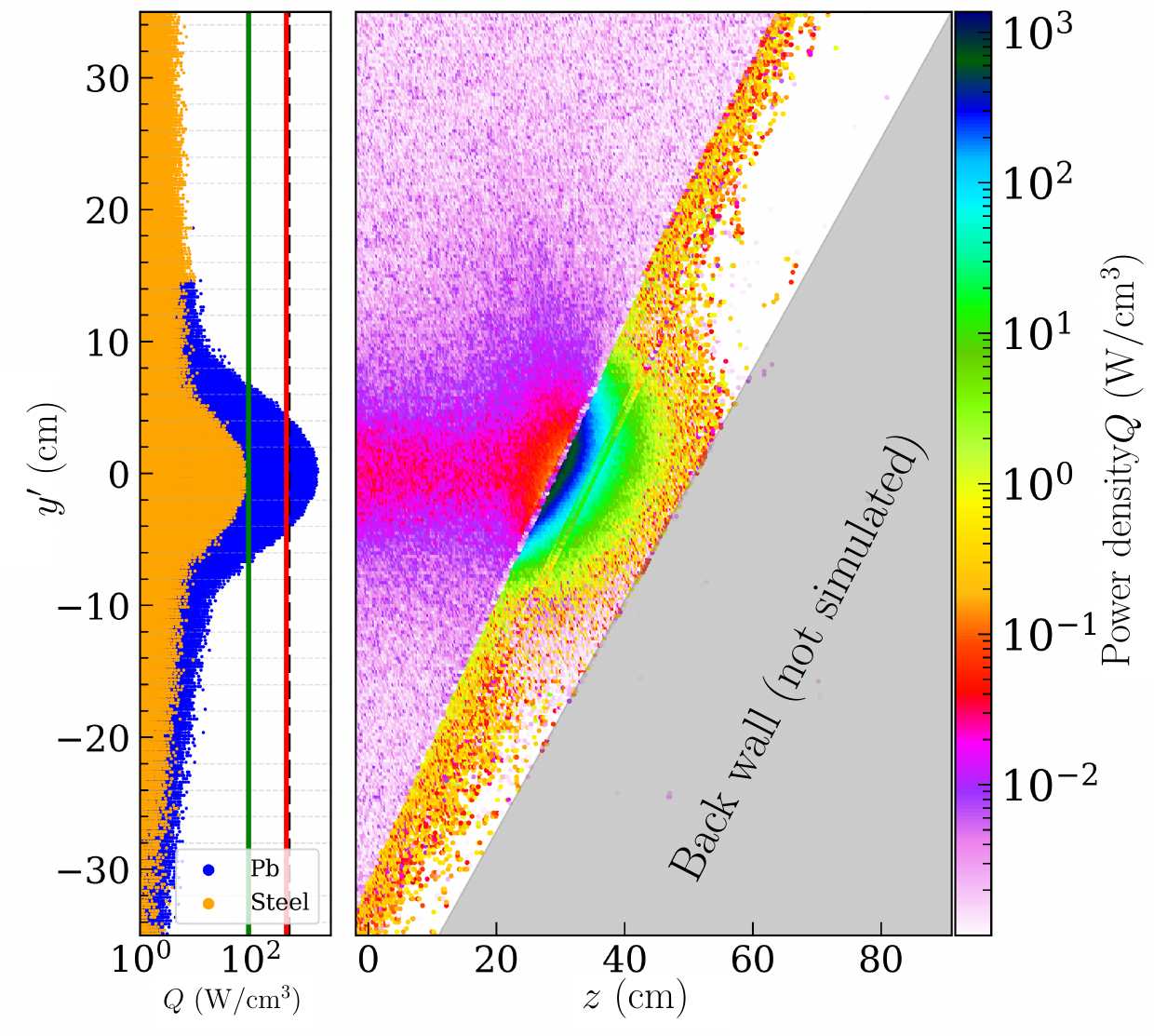}
        \caption{}
        \label{fig:19a}
    \end{subfigure}
    \begin{subfigure}[b]{0.49\linewidth}
        \includegraphics[width=0.85\linewidth]{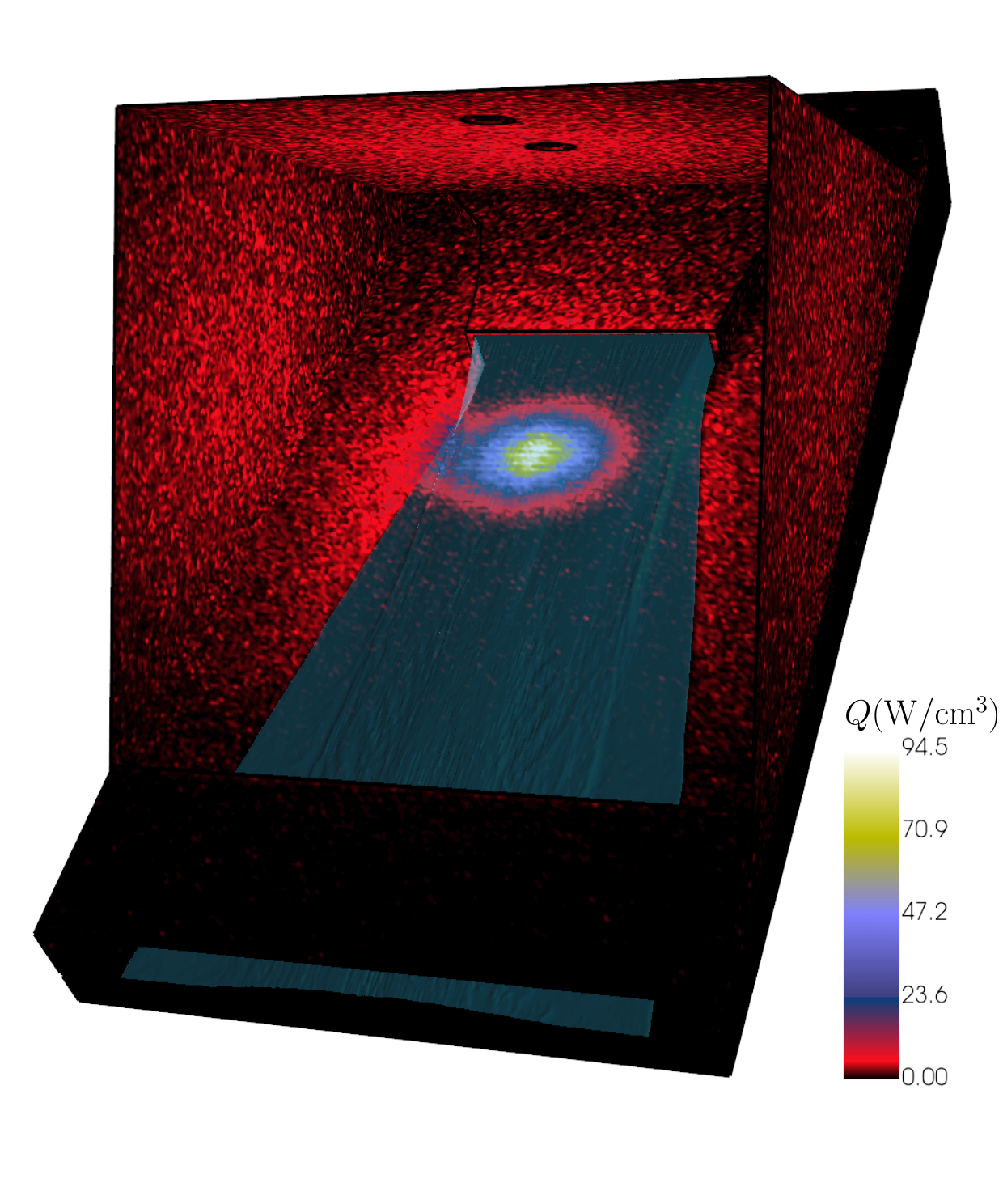}
        \caption{}
        \label{fig:19b}
    \end{subfigure}
    \caption{Compact design with focused liquid-lead curtain for peak power absorption: (a)~power-deposition cut plane and (b)~three-dimensional rendering showing the power density on the steel vessel as a color map.}
    \label{fig:19}
\end{figure}

This optimized configuration creates a more dynamic free flow of liquid lead. Therefore, \textsc{fluka} simulations were performed using this exact fluid geometry to verify the distribution of absorbed power [\cref{fig:19a}]. The results show that the center of the beam spot remains fully contained within the upstream curtain.

\Cref{fig:19a} compares the maximum power density in the steel for the simple and optimized geometries. The optimization reduces the peak by 79.8\%, corresponding to only 20\% of the allowable limit. \Cref{fig:19b} illustrates the energy passing through the free-surface flow before being dissipated in the downstream pool.

\begin{figure}[!htb]
    \centering
    \begin{subfigure}[b]{0.49\linewidth}
        \includegraphics[width=\linewidth]{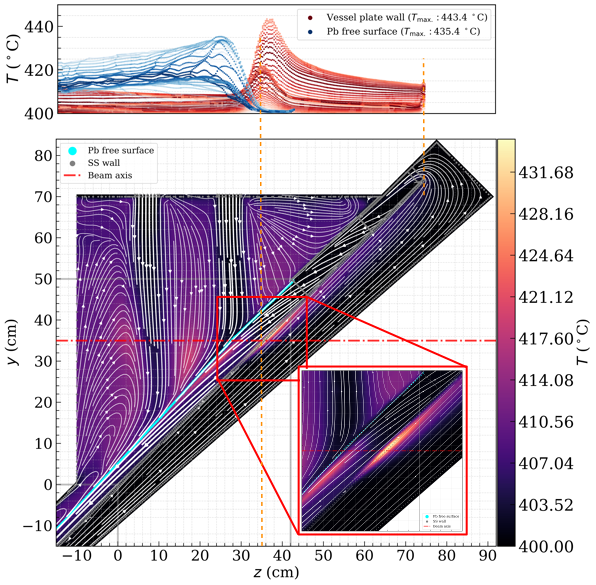}
        \caption{}
        \label{fig:20a}
    \end{subfigure}
    \begin{subfigure}[b]{0.49\linewidth}
        \includegraphics[width=\linewidth]{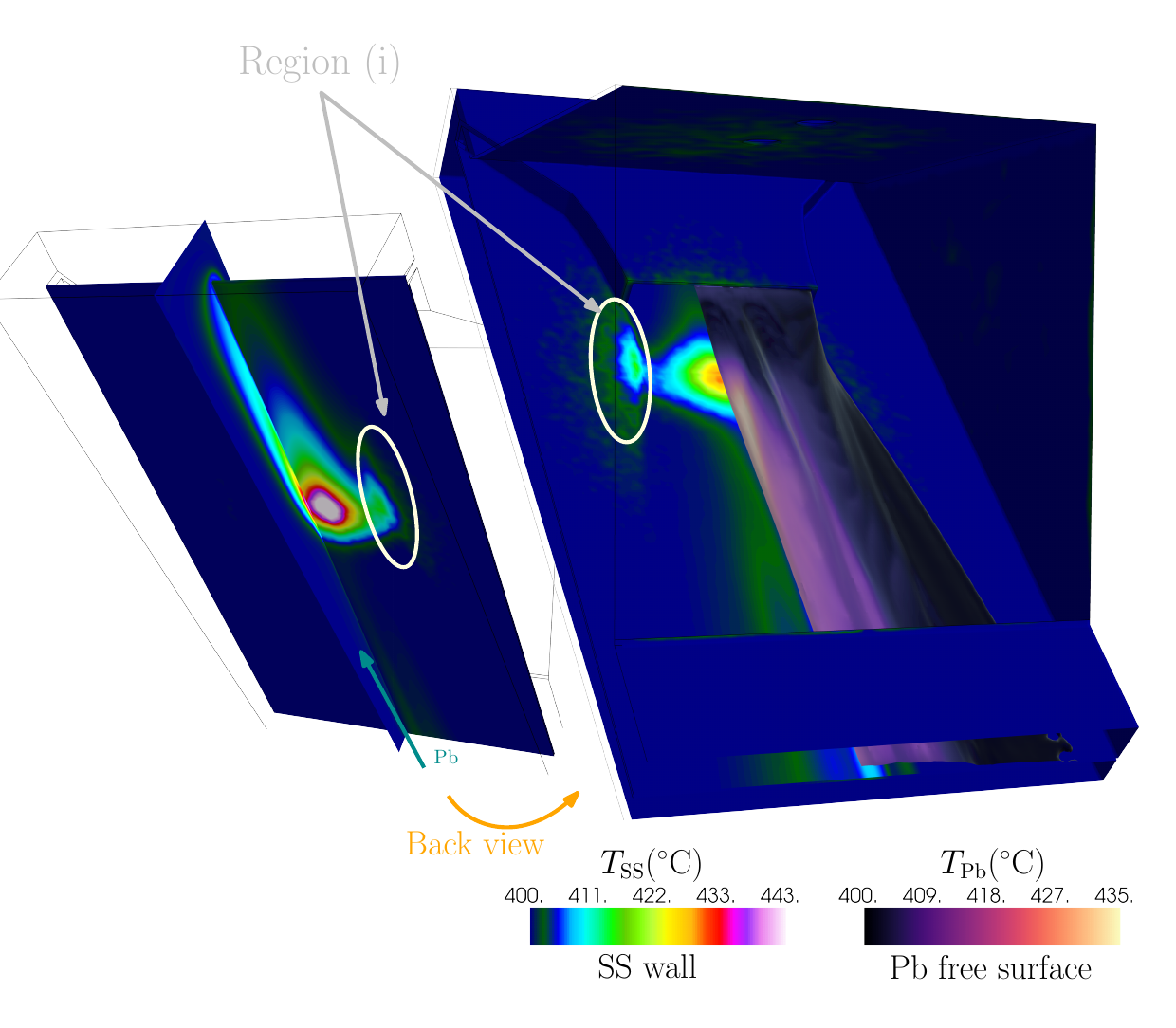}
        \caption{}
        \label{fig:20b}
    \end{subfigure}
    \caption{Compact design with focused liquid-lead curtain: (a)~velocity field and (b)~temperature distribution on the walls and free surface.}
    \label{fig:20}
\end{figure}

The CFD simulations, which were performed under identical boundary and material conditions as the previous ones, reveal the steady-state thermal distribution, as shown in \cref{fig:20}. The maximum steel-wall temperature reaches approximately \SI{443}{\celsius}, but due to convective heat removal to the liquid lead, the interface temperature stabilizes around \SI{435}{\celsius}, consistent with the design threshold for the power density deposited at the Pb-steel interface. 

\cref{fig:20a} shows that the heated region in the argon cover gas (visible in the power-deposition map of \cref{fig:19a}) is strongly distorted by recirculating flow structures. This flow recirculation is induced by the continuous tangential motion of the liquid-lead free surface, which entrains the adjacent argon and promotes convective mixing. As a result, heat deposited in the gas phase does not accumulate locally but is redistributed within the argon volume. This shear-driven mixing prevents the formation of stagnant hot gas pockets near the interaction region and avoids any significant thermal insulation of the Pb-SS interface. 

\cref{fig:20b} further highlights region (i), which corresponds to a power deposition zone that is not covered by liquid lead on both sides. In this area, the steel wall is exposed directly to the photon beam, from the upstream side, while heat removal occurs predominantly through convective cooling by the liquid-lead flow downstream of the wall. Despite this, the resulting wall temperature remains within acceptable limits, stabilizing below the design threshold, and confirming our acceptable peak power on the SS walls. This shows that the optimization of the free-surface to the peak power region provides a reliable solution with an efficient thermal safety.

Compared with the S-slope absorber, the CUSP design achieves similar absorption efficiency (approximately 95\% of total beam power) within a length of only \SI{1}{\meter}, reducing the structural volume by a factor of~5. The absence of backscattering eliminates the need for an upper deflector, while the compact geometry simplifies integration and shielding.

The temperature field shown in \cref{fig:20b} exhibits a narrow, oblique hot streak following the flow trajectory, while the streamlines indicate minor recirculation in the argon medium but not in the lead flow. This configuration promotes efficient convective removal and maintains safley the temperature of the liquid lead below \SI{450}{\celsius}, consistent with the thermal design margins.

The main trade-offs of the CUSP concept are higher local velocities, a larger pressure drop, and potential free-surface instabilities. These are mitigated through the use of smooth contractions and rounded deflectors, and the avoidance of supercritical regimes ($\mathrm{Fr} \gtrsim 1$) in the impact region. The short $45^\circ$ film should be smoothly connected to the pool to prevent splashing or hydraulic jumps. The free-surface stability, assessed under the nominal $\dot{m}$, remained within acceptable limits for the \SI[number-unit-separator=\text{-}]{2}{\centi\meter} case. The total pressure-drop budget for the compact circuit remains comparable to the baseline, as the concept does not require higher flow rates or extended piping.

\subsection{Radiation protection and shielding}

BS emitted from the IPs and absorbed by a dedicated beam dump creates specific radiological hazards. The associated risks must be managed and accounted for during the design phase of the beam absorbers.

\begin{figure}[!htb]
    \centering
    \includegraphics[width=0.6\textwidth]{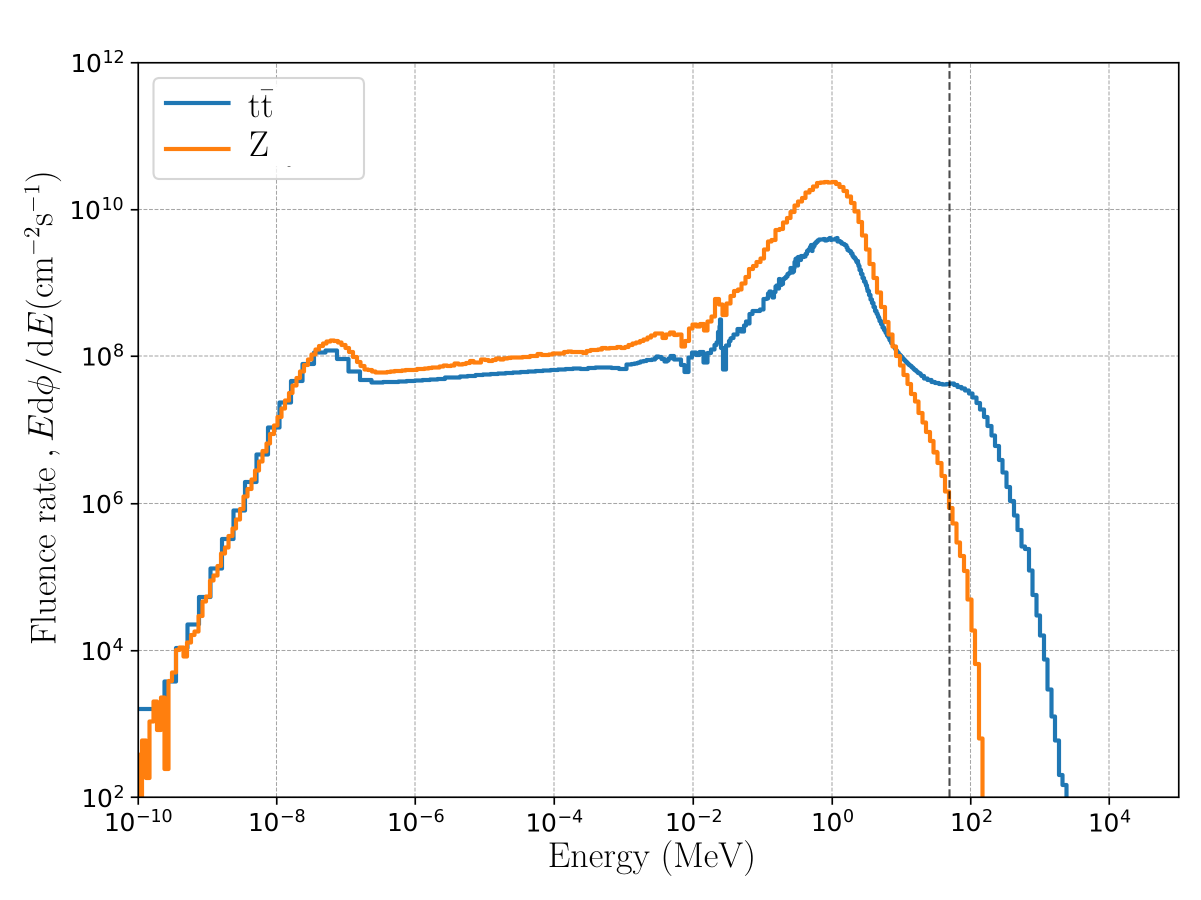}
    \caption{Neutron fluence spectra inside the liquid-Pb absorber for the Z and ${\rm t}\bar{\rm t}$ modes. The dashed vertical line at \SI{35}{\mega\electronvolt} marks the approximate threshold for the onset of fission in Pb isotopes.}
    \label{fig:neutron_spectra}
\end{figure}

The BS spectra vary significantly between the Z and ${\rm t}\bar{\rm t}$ modes, with critical energies of 6.5 and \SI{205}{\mega\electronvolt}, respectively. Although the overall power is higher in Z-pole operation due to the higher beam current, the spectrum in ${\rm t}\bar{\rm t}$ mode extends to higher energies in the \si{\giga\electronvolt} range, which changes the radiological impact in several ways. The photon flux above \SI{7}{\mega\electronvolt} is high enough in both modes to produce photoneutrons in the beam absorber and surrounding materials. These neutrons will undergo nuclear interactions that lead to activation of the absorber and surrounding infrastructure. The neutron spectrum extends up to about \SI{100}{\mega\electronvolt} in Z-pole operation and \SI{1}{\giga\electronvolt} in the ${\rm t}\bar{\rm t}$ mode, with a significant peak at about \SI{1}{\mega\electronvolt} in both cases. \Cref{fig:neutron_spectra} presents the simulated neutron spectra in the lead absorber for the Z and ${\rm t}\bar{\rm t}$ modes. The fission reactions in lead are more relevant in ${\rm t}\bar{\rm t}$ mode, given the higher neutron-energy reach.

The prompt radiation must be shielded to protect the equipment in the immediate vicinity of the dump and to limit activation of these components. In an initial design study, the dump tank was therefore enclosed by a \SI[number-unit-separator=\text{-}]{1}{\meter}-thick concrete shielding layer to provide minimum containment for penetrating neutron radiation and residual gamma radiation. The shielding thickness, material choices, and decay-heat-removal strategies must be optimized to meet radiation-protection requirements for both prompt and residual doses.

\begin{figure}[!htb]
    \centering
    \includegraphics[width=1\textwidth]{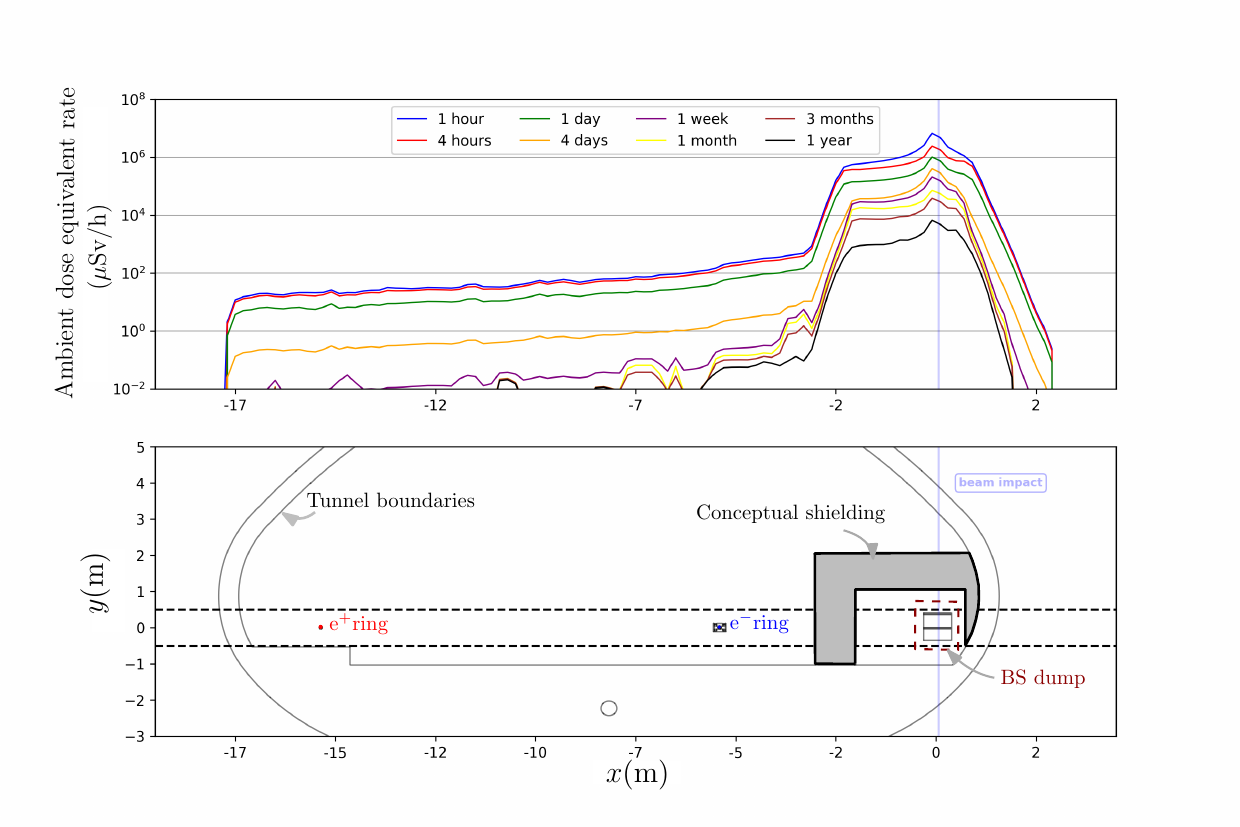}
    \caption{Average residual radiation levels after \SI{1}{yr} of Z-pole operation for different decay times in a transversal plane at the level of the BS dump ($\mathrm{e}^+$ and $\mathrm{e}^-$ positions are symmetric with the IP).}
    \label{fig:dose_rates_BS_dump_z_pole}
\end{figure}

The radiation environment around the BS dump was evaluated for the Z~mode assuming an irradiation of \SI{185}{\day\per yr}. The dose rate inside the dump tank reaches \SI{1}{\sievert\per\hour} after \SI{1}{\hour} of decay time. At a distance of approximately \SI{3}{\meter} from the concrete shielding, residual dose rates decrease below \SI{100}{\micro\sievert\per\hour} (see \Cref{fig:dose_rates_BS_dump_z_pole}). Such levels are within reach of acceptable limits for restricted human access under controlled conditions, provided the shielding is further increased and optimized. After \SI{1}{yr} of decay time, the dose rates inside the dump tank remain of the order of several millisieverts per hour, requiring fully remote operation for any maintenance tasks. The initial study did not include the radiological impact of the circulating liquid lead, protective gas, or cooling circuits in the dump's auxiliary infrastructure.

In addition to the residual dose rate, the activation of the dump components and protective gas, as well as the cooling fluids and air, must also be taken into account.

The BS dump, including the tank, absorber, shielding, and all auxiliary equipment such as pumps, storage containers, and measurement and control units, should be enclosed and equipped with a separate ventilation and cooling circuit. In this way, the system can be radiologically isolated to prevent any negative impact on the rest of the accelerator tunnel and equipment, and to control releases toward the tunnel and ultimately the environment.

The liquid-lead absorber will become highly radioactive. During the technical design phase, a detailed analysis of activation and nuclide inventory is required to quantify the radiological risks associated with maintenance, remote handling, and the eventual disposal of activated components. The potential outgassing of radioisotopes must also be considered. In turn, lead has the advantage of providing effective self-shielding for the gamma emitters it contains. A high level of purity, in particular a low concentration of Bi isotopes, is desirable to limit the production of alpha-emitting isotopes. The overall activity produced in the lead is expected to be almost 2~orders of magnitude higher in ${\rm t}\bar{\rm t}$ operation than in ${\rm Z}$ mode due to the higher-energy particle flux involved.

The current design uses argon under light overpressure in the dump tank as the protective gas; however, argon exhibits the highest initial activity concentration, exceeding \(10^7\)~\si{\becquerel\per\gram}, and this remains above \(10^6\)~\si{\becquerel\per\gram} even after several days of cooling. Production of $^{41}$Ar will represent an additional source of gamma radiation that could affect the dose rates of the circuits and the storage container. In contrast, nitrogen and helium exhibit lower activity concentrations with much shorter-lived isotopes, with that of nitrogen dropping below \(10^6\)~\si{\becquerel\per\gram} within \SI{1}{\day} and below \(10^5\)~\si{\becquerel\per\gram} after a year. The radioisotope composition differs significantly between the different support gases. The technical design of the system must specify how to detect and contain gas leaks and how to implement a safe gas-cleaning system. A more detailed study and comparison to select the appropriate protective gas must be performed, and releases from the gas system must be quantified and evaluated for their radiological impact.

The lead will remain radioactive and, after the operation of FCC-ee, will become radioactive waste. Considerable costs will be involved in the conditioning and disposal of this lead. Dismantling and conditioning procedures must be included in the technical design of the beam dumps.

A complete risk analysis must be performed for the dump system---including preventive and corrective maintenance tasks and potential incident and failure scenarios---to ensure that the technical design incorporates the required prevention, containment, and mitigation functions.

\subsection{Engineering aspects and integration of the BS dump}

\begin{figure}[!htb]
    \centering
    \includegraphics[width=1\textwidth]{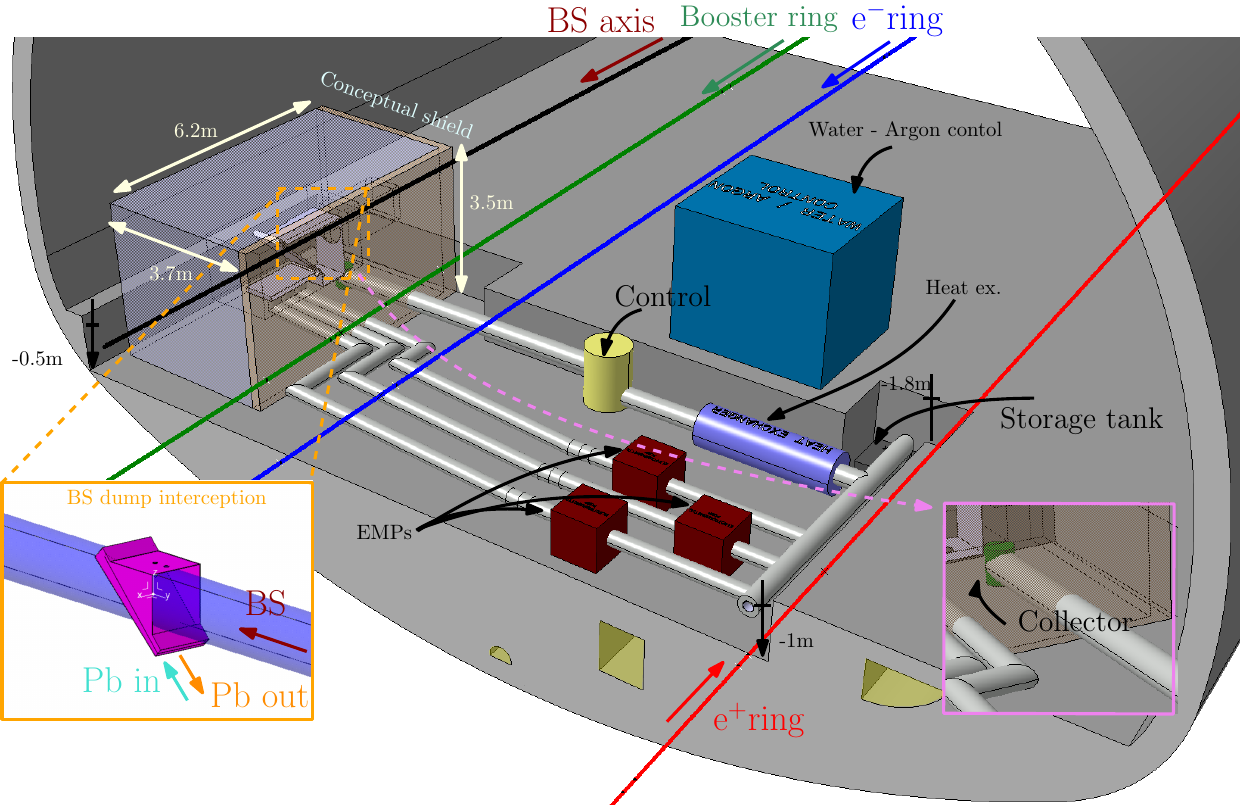}
    \caption{Conceptual integration of the FCC-ee BS dump and hydraulic system in the tunnel. The left-hand inset shows the geometry by which the liquid-lead absorber intercepts the BS photons.}
    \label{fig:BSdumpIntegration}
\end{figure}

\Cref{fig:BSdumpIntegration} illustrates the conceptual integration of the BS dump and its associated hydraulic system into the FCC-ee tunnel environment, also showing the relative positions of the booster, electron, and positron rings. The absorber position for the CUSP concept is shown in the inset.

The liquid lead enters the absorber at temperature of approximately \SI{400}{\celsius} and exits via an outlet connected to a collector after absorbing the photon energy. After the absorber, the liquid lead flows into a collector vessel located downstream. This vessel collects the heated lead, promotes temperature homogenization, and provides an intermediate containment volume. Flow circulation is achieved using three electromagnetic pumps operating in parallel. The heated liquid is transferred through a heat exchanger, where energy is removed and dissipated into a secondary water-based cooling loop. The secondary system operates under argon cover gas to limit oxidation and to maintain stable thermochemical conditions. A storage tank with a capacity of approximately \SI{1.8}{\meter\cubed} is included for liquid-metal inventory control, filling, and thermal expansion.

The absorber is installed inside a shielded enclosure, which occupies a volume of approximately $6.2 \times 3.7 \times \SI{3.5}{\meter\cubed}$, providing radiation protection as discussed before. Thermal insulation and mechanical supports reduce heat losses and maintain alignment stability during operation. The pumps and heat-exchanger units are positioned roughly \SI{1}{\meter} below the beam plane.

\section{Conclusions}\label{sec:5}

This study has drawn attention to the importance of the geometric configuration in optimizing liquid-lead absorbers for BS absorption in FCC-ee. Within the practical hydraulic constraint of a maximum mass flow rate, and given the beam power and size, two viable concepts were analyzed.

A suitably inclined slope---the double-slope design concept---can achieve the required effective thickness while operating within feasible hydraulic limits. Using an optimized curved profile provides clear advantages over a linear slope: it distributes the photon interaction over a larger area, reduces the peak power density, and mitigates localized hot-spot formation. A thermal steady state is reached well below the \SI{500}{\celsius} operational limit, with peak temperatures safely below this threshold, and backscattering is fully absorbed with an upper flow.

A more compact absorber, the CUSP concept, becomes feasible when the initial power spike is intercepted by a thin free-surface curtain, followed by a lead pool that absorbs the residual shower. Despite the inherently small curtain thickness achievable under the mass flow constraint, the CUSP geometry effectively reshapes the photon interaction path, yielding absorption efficiencies comparable to the long-slope concept but within a total length of approximately \SI{1}{\meter}. The thermal field is more uniformly distributed, peak wall temperatures remain below \SI{450}{\celsius}, and integration requirements are significantly reduced relative to a \SI[number-unit-separator=\text{-}]{5}{\meter}-long slope.

Overall, the results show that both concepts can satisfy FCC-ee BS absorption requirements with complementary strengths: the sloped-flow concept maximizes absorption via its extended geometry, while the CUSP concept achieves similar performance within a much more compact footprint. Within the simulated operating envelope, no thermal–hydrodynamic feedback between the heated argon, the steel wall, and the liquid-lead free surface is observed, indicating stable two-phase behavior under nominal operating conditions. Owing to the large thermal inertia of the circulating liquid lead, the system exhibits low sensitivity to short-term beam-power transients, as hydrodynamic steady state is established significantly faster than thermal equilibrium. The present predictions carry conservative estimated uncertainties of $\mathcal{O}(5\%)$ in absorbed power and $\mathcal{O}(10\%)$ in local temperatures, indicating the importance of future experimental validation and prototype testing.

\appendix

\section{\label{app:A}Validation of the numerical fluid-dynamic model}

\begin{figure}[!htb]
     \centering
     \begin{subfigure}[b]{0.49\linewidth}
         \centering
         \includegraphics[width=\textwidth]{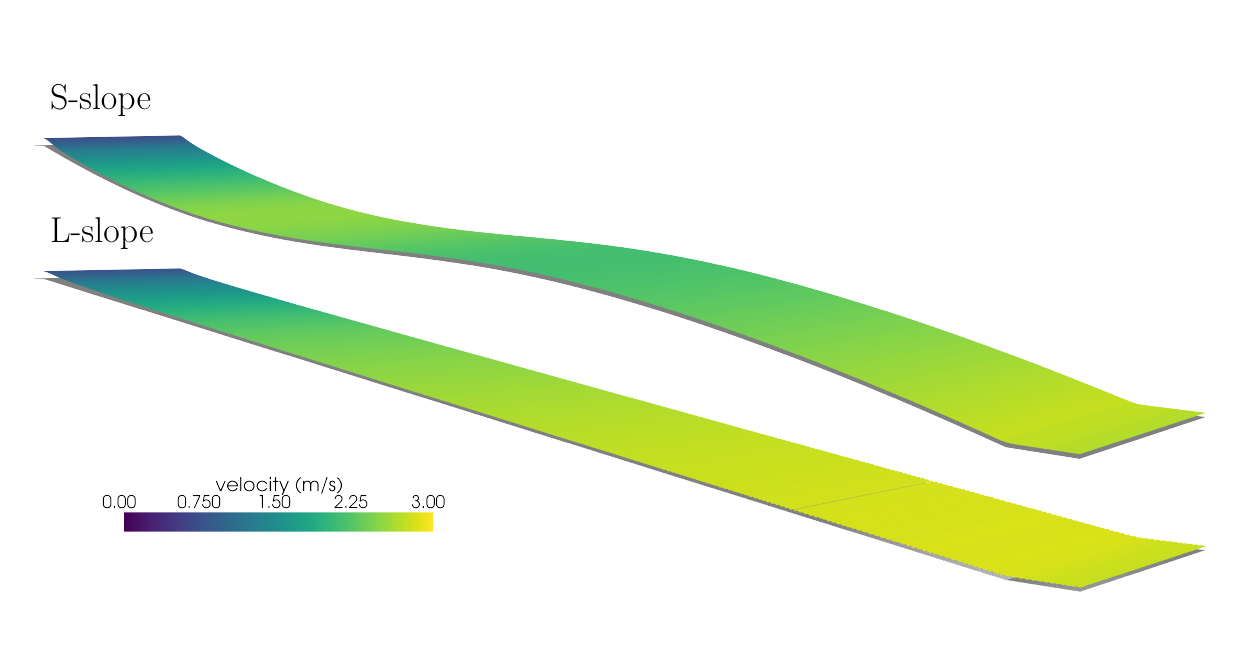}
         \caption{}
         \label{fig:4a}
     \end{subfigure}
     \hfill
     \begin{subfigure}[b]{0.49\linewidth}
         \centering
         \includegraphics[width=\textwidth]{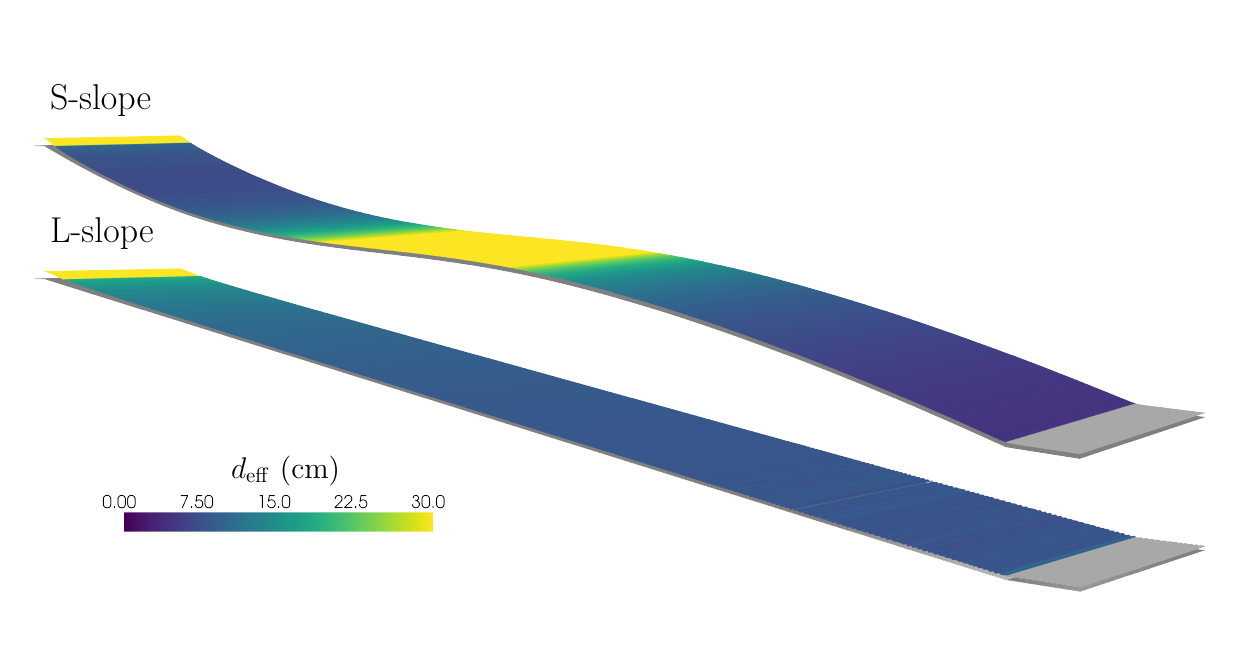}
         \caption{}
         \label{fig:4b}
     \end{subfigure}
     \caption{Two-dimensional simulation contours of (a)~velocity magnitude and (b)~effective thickness $d_{\mathrm{eff}}$ on the free surface of the liquid lead for the L- and S-slope geometries.}
     \label{fig:4}

\end{figure}

To validate the numerical fluid-dynamic model, a one-dimensional hydraulic calculation of the free-surface profile was performed. In these calculations, the flow is assumed to be steady, incompressible, fully turbulent, and gradually varied in a rectangular channel of width $w$. For a prescribed mass flow rate $\dot m$, the local film thickness $h(x)$ determines the cross-sectional area $A = w h$ and bulk velocity $v = \dot m / (\rho A)$. The total head is written as
\begin{equation}
  H(x) = z_b(x) + h(x) + \frac{v(x)^2}{2g},
\end{equation}
and its streamwise variation includes frictional and geometric losses. The friction factor is obtained from the explicit three-parameter Colebrook--White approximation~\cite{Bellos2018a}, which depends on the local Reynolds number and equivalent roughness. The free surface is then advanced stepwise along the channel by updating the head and enforcing mass conservation,
\begin{equation}
  h = \frac{\dot m}{\rho\,w\,v}.
\end{equation}
The resulting free-surface elevation $z_s(x) = z_b(x) + h(x)$ provides a one-dimensional reference profile for comparison with the CFD predictions. \Cref{fig:4} shows the corresponding velocity and effective thickness.

\begin{figure}[!htb]
     \centering
         \includegraphics[width=0.8\textwidth]{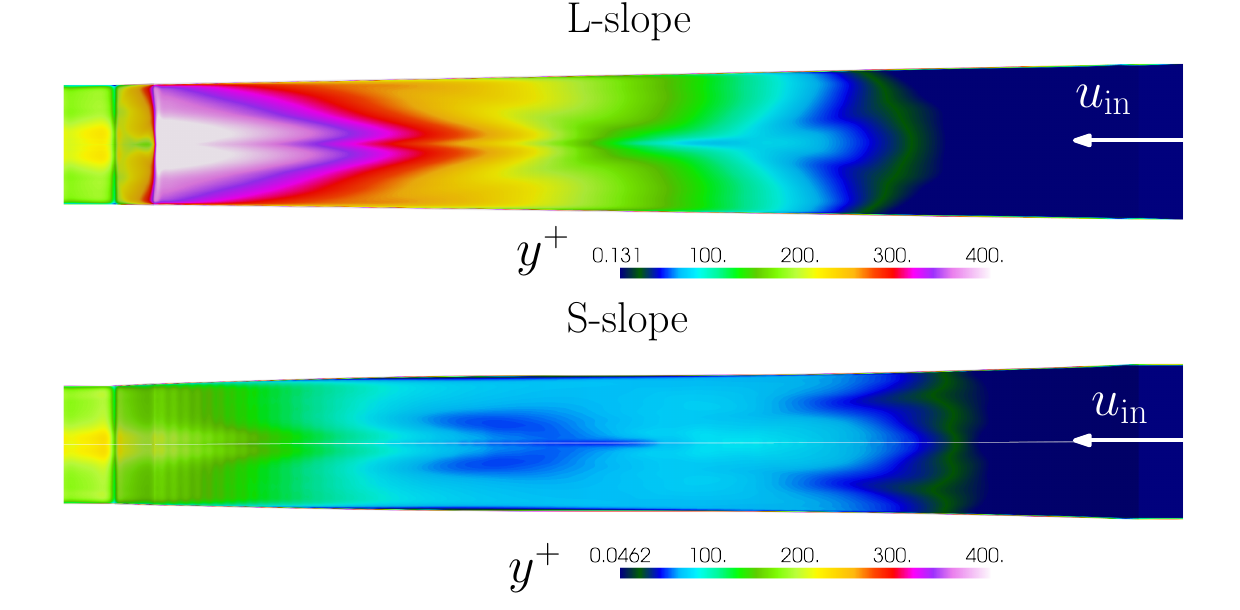}
        \caption{Distributions of $y^{+}$ from the simulations of the single-slope configurations.}
        \label{fig:ap2}
\end{figure}

\Cref{fig:ap2} shows the resulting $y^{+}$ distributions along the walls for the single-slope configurations. The values fall within the desired range for the wall-treatment approach adopted in this study, indicating adequate near-wall resolution for the turbulence model. Only a small region near the end of the L-slope exhibits elevated $y^{+}$ values; however, this occurs in a zone where the local free-surface velocity reaches about \SI{3}{\meter\per\second}, a condition that does not occur in the full-design configurations (double-slope and CUSP). Consequently, this local deviation does not affect the validity of the near-wall modeling for the geometries considered in the final designs. For the double-slope and CUSP cases, the near-wall cell size and growth rate are kept unchanged, so the resulting values remain within the same target range throughout the domain.

\section{\label{app:B}Solution details}

Each time step was considered to have converged if the absolute residuals reduced to $10^{-3}$ or the relative residuals decreased by 3~orders of magnitude for the continuity and momentum equations; however, for the energy equation, convergence was taken as having been achieved when the absolute residuals reduced to $10^{-6}$. The wall temperatures at different sections were monitored throughout the calculations, and the case was advanced until the temperature variations reached a semisteady state, defined as a variation of 5\% over time.

An adaptive time-stepping method was applied to maintain a Courant number of 0.3, ensuring numerical stability and accuracy. The Courant number is defined as
\begin{equation}
C = \frac{u \Delta t}{\Delta x},
\end{equation}
where $u$ is the flow velocity, $\Delta t$ is the time step, and $\Delta x$ is the grid spacing. The time step was adjusted automatically, within the range $10^{-6}$ to $10^{-3}$~s, to keep the Courant number below 0.3 and ensure good temporal resolution.

The pressure-implicit with splitting of operators (PISO) scheme was used for pressure-velocity coupling. This scheme is well suited to transient simulations and multiphase unsteady flows. The pressure-based solver in \textsc{ansys} \textsc{fluent} was employed, and the spatial discretization for pressure was set to second-order accuracy. A VoF method with level-set improvement was used to ensure good interface accuracy and minimize the required number of cells, due to the geometric reconstruction used in the method~\cite{OLSSON2005225, Candido2021}.


\bibliography{apssamp-edited}

\end{document}